%% file: OrionGEMS.tex
\documentclass{aa}

\usepackage{deluxetable}
\usepackage{rotating}

\newcommand\arcdeg{\mbox{$^\circ$}\xspace}  

\newcommand{\cmq}{cm{$^{-3}$}}

\newcommand{\kms}{km~s{$^{-1}$}}

\newcommand{\Ha}{\mbox{H$\alpha$}}
\newcommand{\sii}{[S~{\sc ii]}}
\newcommand{\Feii}{[Fe~{\sc ii}]}

\newcommand{\oi}{[O~{\sc i}]}
\newcommand{\nii}{[N~{\sc ii}]}

\newcommand{\Vlsr}{V$_{LSR}$}

\newcommand{\um}{$\mu$m}
\newcommand{\mum}{\ensuremath{\mu \mathrm{m}}}

\newcommand{\hh}{\ensuremath{\textrm{H}_{2}}}			

\titlerunning{Orion GeMS GSAOI}

\begin{document}

\title{The Orion Fingers:   Near-IR Adaptive Optics Imaging of
an Explosive Protostellar Outflow}

\author{
John Bally{\cu},
Adam Ginsburg{\eso},
Devin Silvia{\msu},
and
Allison Youngblood{\cu}
	    }
\authorrunning{Bally et al}

\newcommand{\cu}{$^{1}$}
\newcommand{\eso}{$^{2}$}
\newcommand{\msu}{$^{3}$}
	    
\institute{{\cu}{\it{
      Department of Astrophysical and Planetary Sciences,\\
      University of Colorado, UCB 389, \\
      Boulder, CO 80309}      \\
      \email{John.Bally@colorado.edu}
      }\\
{\eso}{\it{
     ESO Headquarters\\
     Karl-Schwarzschild-Str. 2\\
     85748 Garching bei M{\"u}nchen,
     Germany}      \\
}\\
{\msu}{\it{
      NSF Astronomy and Astrophysics Postdoctoral Fellow, \\
      Department of Physics and Astronomy,\\
      Michigan State University, \\
      East Lansing, MI 48824}\\
}
    }

\keywords{ISM: - molecular clouds --
ISM: - shocks, outflows
ISM: individual -- Orion Nebula, OMC1
stars: formation -- }

\abstract
{}
{Adaptive optics images are used to test the hypothesis
that the explosive  BN/KL outflow from the Orion OMC1 cloud 
core was powered by the dynamical decay of a non-hierarchical
system of massive stars.  }
{
Narrow-band \hh ,  \Feii, and broad-band K$_s$ obtained with the Gemini 
South multi-conjugate adaptive optics (AO) system GeMS and near-infrared 
imager GSAOI are presented.    The images reach resolutions of  0.08 to 
0.10\arcsec, close to the 0.07\arcsec\ diffraction limit of the 8-meter 
telescope at 2.12 \mum.  Comparison with previous AO-assisted observations of 
sub-fields and other ground-based observations enable measurements of proper 
motions and the investigation of morphological changes in  \hh\ and \Feii\ features 
with  unprecedented precision.   The images are compared with numerical simulations of 
compact, high-density clumps moving $\sim 10^3$ times their own
diameter through a lower density medium at Mach $10^3$. 
}
{
Several sub-arcsecond \hh\ features and many \Feii\ `fingertips' on the 
projected outskirts of the flow show 
proper motions of $\sim$300 \kms.    High-velocity,  
sub-arcsecond   \hh\  knots (`bullets')  
are seen as far as 140\arcsec\ from their suspected ejection site.    
If these knots propagated through the dense  Orion A 
cloud, their survival  sets  a lower bound on their densities of order $10^7$ \cmq ,  
consistent with an origin within a few au of a massive star and accelerated
by a final multi-body dynamic encounter  that ejected  
the BN  object and radio source I from OMC1 about 500 years ago.  
}
{
Over 120 high-velocity bow-shocks propagating in nearly all 
directions from the OMC1 cloud core provide evidence for an explosive 
origin for the BN/KL outflow triggered by the dynamic decay of a non-hierarchical
system of massive stars.   Such events may be linked to the origin of 
runaway, massive stars. 
}

\maketitle

\section{Introduction} 

The BN/KL region behind the Orion Nebula, located at a distance of about
414 pc \citep{Menten2007},  contains a spectacular,  wide opening-angle, 
arcminute-scale outflow emerging from the OMC1 cloud core.  The flow is 
traced by the millimeter and sub-millimeter emission lines of molecules 
such as CO, CS, SO, SO$_2$, and HCN  that exhibit  broad  
($>$ 100 km s$^{-1}$)  emission line wings 
\citep{KwanScoville1976, WisemanHo1996, FuruyaShinnaga2009},  
high-velocity OH, H$_2$O,  and SiO maser  emission \citep{Genzel1981,Greenhill1998}, 
and bright shock-excited `fingers' of  \hh\   and `fingertips' of 1.64 \um\  \Feii\   
emission \citep{AllenBurton93, Colgan2007, Nissen2007, 
LeeBurton2000, Bally2011}.    The OMC1 outflow has a southeast 
(red-shifted) to  northwest (blue-shifted) axis and contains  at least 8 M$_{\odot}$ of  
accelerated gas with a median velocity of about  20 km s$^{-1}$.     Interferometric 
CO images, H$_2$O and the 18 km~s$^{-1}$  SiO masers,   and dense-gas  
tracers such as  thermal SiO emission reveal a  smaller (8\arcsec\  long) 
and younger  ($\sim$ 200  year old) outflow along a  northeast-southwest axis 
emerging  from radio source I   orthogonal to the arc-minute-scale  CO outflow  
\citep{BeutherNissen2008,Plambeck2009}.    
The  momentum and kinetic   energy content of these flows is at least  
$160$ M$_{\odot}$ km s$^{-1}$ and  $4 \times 10^{46}$  ergs  \citep{Snell1984} to  
$4 \times 10^{47}$ ergs  \citep{KwanScoville1976}.     \citet{Zapata2009}  presented 
a CO J = 2--1 interferometric study and found a dynamic age of  about $500$ 
years for the larger OMC1 outflow.   They noted that its impulsive nature and 
its  structure is  different from accretion-disk  powered  jets and collimated 
protostellar outflows,  and that it  originated several arcseconds 
north of the  OMC1 hot-core.       

Radio-frequency astrometry has shown that
the three  radio-emitting stars in OMC1, sources BN, I, and 
source n, have  proper motions  of 25, 13, and 26 km s$^{-1}$ 
away from a region less than 500 au in diameter from which they were ejected 
about  500 years ago \citep{Rodriguez2005,Gomez2005,Gomez2008}.    Although
the motions of massive stars BN and source I have been confirmed by many 
independent measurements of the positions of maser spots and radio continuum 
emission,  the motion of source n has been questioned \citep{Goddi2011}.  
Determinations of its  proper  motion may be influenced by intensity 
variations of its bipolar radio nebula which may shift source n's  emission centroid. 
  
\citet{BallyZinnecker2005} proposed that the  BN/KL explosion may have
been triggered by a massive star merger or  the dynamical rearrangement 
of a nonhierarchical system of massive stars into a hierarchical system 
that resulted in their ejection from the OMC1 core as high-velocity, runaway stars.  
In this model,  the disruption and ejection of 
circumstellar  disks and envelopes  produced  the BN/KL outflow.  The 
momentum and kinetic energy of the outflow and ejected  stars came from 
the release of gravitational  binding energy of a compact binary  formed  
by the dynamic interaction of  three of more stars 
\citep{PovedaRuizAllen1967,Gualandris2004,Perets2012,ReipurthMikkola2010,ReipurthMikkola2012}.  
 
\citet{Zapata2009},  \citet{Bally2011}, and \citet{Goddi2011} found supporting 
evidence for the  dynamical  rearrangement model.    In this scenario, 
the final N-body encounter would have resulted in the formation 
of a compact,  au-scale binary,   most likely source I , and the ejection of 
radio source  I and BN, and possibly source n.    However, it is still possible 
that additional highly-embedded IR sources will be found in dense cores such
as the 1 millimeter source, SMA1  \citep{BeutherNissen2008}.  Some of these
might contain massive stars.   After all, radio source I has not yet been directly
detected at any infrared wavelength.

Over 30\% of massive stars are ejected from their birthplaces at high velocities
\citep{GiesBolton1986}.   A large fraction of such runaway stars are thought
to have been produced by the {\it dynamical ejection scenario}
\citep{PovedaRuizAllen1967}.    For example, a  dynamical  interaction
is thought to have ejected the classic runaway stars $\mu$ Columbae and
AE Aurigae with velocities of 108 \kms\  and 113 \kms ,  respectively,  in opposite
directions about 2.5$\pm$0.05 Myr ago \citep{Hoogerwerf2000, Hoogerwerf2001,
Gualandris2004}.    This event left behind the colliding-wind X-ray binary,
$\iota$ Orionis, which is the brightest and most massive member of the
$\sim$5 Myr old NGC~1980 cluster located about 30\arcmin\ south of the
Orion Nebula in the Orion A molecular cloud \citep{AlvesBouy2012,Bouy2014}.  

Proper motion  measurements show that the  fastest components in the  OMC1 
fingers have a  dynamic age of about 500  years \citep{Bally2011}.     However,   
\citet{Wu2014arXiv1} used ALMA observations to argue for a somewhat older
dynamic age of $<$1,300 years for the BN/KL outflow.    \citet{Bally2011} found 
that  the \hh\  and \Feii\  finger proper motions trace-back to within a 
few arcseconds of   J2000 = 05:35:14.5, $-$05:22:23 located between the current 
locations of  radio sources I and BN.     \citet{Gomez2008} found that in the Orion 
reference frame, the proper motions of radio sources I, BN, and n intersected 
within one arcsecond ($<$ 400 au) of  J2000 = 05:35:14.360, $-$05:22:28.70  
about 500 years ago, a few arcseconds from the apparent point of outflow origin 
determined by tracing back the near-infrared proper motions of the fast ejecta. 
 
\citet{Tan2004} proposed an alternative scenario in which the BN object was ejected 
about 4,000 years ago by a dynamic interaction in the Trapezium cluster located 
in the center of the Orion Nebula.   In this scenario,  the OMC1 explosion was
triggered by the  serendipitous passage of the BN object through the OMC1 core.   
Although  requiring a highly unlikely close encounter of  BN with source I in the 
OMC1 core,  \citet{Tan2012} show that the parameters of the Trapezium and 
BN are compatible with  this scenario.
 
Though rare, the explosive outflow morphology of the  OMC1 BN/KL 
outflow is not unique; other possible  examples include G34.25+0.16 in the inner
Galaxy \citep{Cyganowski2008},  source G in W49 \citep{Smith2009}, 
IRAS 05506+2414  \citep{Sahai2008}, DR21 \citep{Zapata2013}
and possibly the  low-luminosity source
of the molecular hydrogen outflow MHO 2714 (GGD 34) in NGC 7129
\citep{Eisloffel2000}.   However,  Orion BN/KL 
is the nearest and least  obscured,  and thus most accessible for high-resolution 
studies.   Here, we present 0.08 to 0.1 arc second resolution images of the 
entire OMC1 outflow complex in the 1.64 \um\ \Feii\  and 2.12 \um\ \hh\  narrow-band filters 
and a broad-band K$_s$ filter obtained with adaptive optics (AO) on the Gemini South  8 
meter telescope.    This data is combined with  older AO-assisted observations
obtained on Gemini North and natural seeing-limited images acquired with a variety
of other telescope to re-measure proper motions in parts of the BN/KL outflow.

\section{Observations}

\subsection{Gemini South GeMS}

The Gemini Multi-conjugate adaptive optics System (GeMS) at the Gemini South telescope
on Cerro Pachon is the first and only sodium-based multi-laser guide star (LGS) adaptive
optics system \citep{Rigaut2014,Rigaut2012,Neichel2014,Neichel2013,dOrgeville2012}.
GeMS works with a LGS constellation of 5-spots:  4 of the LGS spots are at the corners of a
60\arcsec\ square, with the 5-th positioned in the center.  The Adaptive Optics (AO)  bench
called Canopus is mounted on one of the f/16 Cassegrain ports.    Gemini South 
Adaptive Optics Imager (GSAOI) is a wide-field 4096 by 4096 pixel 
(85\arcsec\ by 85\arcsec\ field of view) 
camera designed to work at the diffraction limit of the 8-meter telescope in the near-infrared.
Three 85\arcsec\  diameter fields were observed in OMC1 between 30 December 2012 and
28 February 2013 using GSAOI.  Observations of each field were obtained through 1\%
bandpass narrow-band filters centered on the 1.644 $\mu$m [FeII] and 2.122 $\mu$m
H$_2$ emission lines and the broad-band K$_s$ filter.  
The corrected images have FWHM diameters of 0.08\arcsec\ to  0.1\arcsec , providing the 
highest angular resolution images of the BN/KL outflow ever obtained in the near-IR. 

Each field was imaged in each filter with a 5 point dither pattern to fill-in gaps between the 
four 2048 by 2048 pixel arrays in GSAOI.    Exposure times were 43.4 seconds per exposure 
for \hh, 43.0 for [Fe II], and 15.0 for Ks.  10 exposures were taken in each filter for a total of 
430 seconds on-source in the narrow-band filters and 150 seconds in the continuum filter.

Data were processed with the Gemini GSAOI pipeline.  However, additional astrometric
corrections were required.   Individual exposures were first registered to the
\citet{Muench2002a} catalog sources to acquire a world coordinate system with
RMS pointing error $\sim0.1$\arcsec .     A new catalog of relative star positions
was generated from a preliminary aligned and co-added stack of images and used
to derive a distortion map for GSAOI.    The individual distortion corrected images were 
re-aligned and co-added to form the final mosaic in each filter.  

\subsection{Gemini North Altair}

Between 2007 and 2009, the Gemini North 8 meter telescope was used to observe 
the OMC1 region on three occasions using the Altair AO system with the NIRI near-IR 
camera through  1\% narrow-band \Feii , \hh , and broad-band K$_s$ filters.    
NIRI was used in a configuration which
delivers a  40\arcsec\ field of view.  In the narrow-band filters, a dithered set of five to ten 
30 second  exposures were obtained.    A similar set of 10 second exposures were
acquired in the K$_s$ filter.     In 2007, only the `\hh\  fingers' region was observed as part
of the commissioning of the Altair AO system using the NIRI camera.   During 2008 and 2009,
we intended to image a 3 $\times$ 4 point grid to cover the full extent of the BN/KL outflow.
However, only 5 and 8 fields were actually observed in 2008 and 2009, respectively.    Only
the  `\hh\  fingers'  field was observed during each of the three years.   A summary of the 
observations is given in Table~1.   The angular resolution of the NIRI images ranges from 0.1 
to 0.2\arcsec . 

In the analysis presented here, the NIRI images were registered to the final GSAOI mosaic
using IRAF tasks GEOMAP and GEOTRAN applied to unsaturated field stars.    
Proper motions were determined by marking  the photocenters of features on the multi-epoch
images.     Images of the OMC1 outflow obtained with the 
Subaru 8 meter telescope on MJD = 51484 in 1999 \citep{Kaifu2000},
the Apache Point Observatory 3.5 meter on MJD = 53331 in 2004  \citep{Bally2011}
were also used.    The interval between the Subaru and Gemini south observations 
is 4,839 days.   

\section{Results}

The 2013 epoch GSAOI images presented here reach the near-IR diffraction limit of an 
8-meter telescope and provide the sharpest views obtained thus far of the 
entire  OMC1 BN/KL  outflow.      The combined \hh , \Feii , and K$_s$ color 
image (Figure \ref{fig1}) shows  \hh\ fingers 
tipped with \Feii\  emission extending from about 30\arcsec\ to 140\arcsec\ from the OMC1 core.  
For the analysis of dynamic ages for various features, we assume that all features originated
from J2000 = 05:35:14.360, $-$05:22:28.70 (marked with a cyan cross in Figure \ref{fig2}),
the suspected location from which the BN object 
and radio source I  were ejected about 500 years ago  \citep{Gomez2008} and within a 
few arc seconds of the suspected point of origin of the fingers determined from the
intersection point of the proper motion vectors \citep{Bally2011}.   
Figure \ref{fig2} shows the \hh\ image with the location of BN, I, an n shown along with
their proper motions.

The two brightest \Feii\  bow shocks  correspond to the Herbig-Haro objects 
HH~201  and HH~210 located 60\arcsec\ northwest and  113\arcsec\  north of  OMC1
\citep{Gull1973,MunchTaylor1974,Canto1980,AxonTaylor1984}.   
These shocks are visible on ground-based and Hubble Space Telescope images in
\oi  ,  H$\alpha$,  \nii\, and \sii .  However, they  only exhibit  faint  \hh\ emission
\citep{Graham2003}, indicating that
they lie in the mostly atomic,  photon-dominated region (PDR) located between the Orion
Nebula's ionization front and the background Orion A molecular cloud and OMC1 cloud core.    
The most prominent  \hh\ fingers consist of  multiple \Feii\  finger-tips trailed by \hh\ wakes
with an  orientation of PA $\sim$ 340\arcdeg\ to 350\arcdeg  .   This chain of shocks and
wakes extends  from about 50\arcsec\  to 135\arcsec\ from the ejection center.  It
consists of  at least a dozen  nested \hh\ bow shocks tipped  with \Feii\ emission regions.      
The brightest  \Feii\  features are visible in the visual-wavelength  emission lines 
commonly seen in Herbig-Haro objects such as \oi , \sii , \nii , and \Ha\  and have
been designated as  HH~205 through 209.  These HH objects 
are associated with the tips of the chain of \hh\  wakes propagating toward position angle
(PA) $\sim$  340\arcdeg\ to 350\arcdeg .   

More than 120 distinct wakes are visible in the 2.12 \um\  \hh\  images
that exhibit nearly parallel walls and large proper motions along their axes
\citep{Bally2011}.        Figure \ref{fig3}  shows a median filtered version of the 
2013 GSAOI \hh\ mosaic  created by convolving the  image with a 51 pixel 
kernel (1\arcsec ; each pixel is   0.02\arcsec\  on a side) using the 
IRAF function MEDIAN, and subtracting  the result from the original image.   
Vectors were drawn from the suspected ejection site of radio sources BN and 
I (the coordinates are given  above) to each \hh\ or \Feii\ fingertip.  
The dashed vector near  the top marks a chain of  \hh\ knots and bow shocks with 
proper motions nearly orthogonal to the  northern fingers;  this feature traces 
another  flow originating east of the imaged field.
This flow is also seen faintly at visual wavelengths.    
The \hh\  emission becomes too confused within $\sim$30\arcsec\  of the suspected 
ejection location due to multiple overlapping features.     The natural seeing-limited 
1999 epoch Subaru telescope image from \citet{Kaifu2000} and the 2005 epoch 
image from \citet{Bally2011} were used to  trace a few additional \hh\  fingers 
beyond the boundaries of the GSAOI image in the southwest portion of Figure \ref{fig3}.    

In the northwestern  part of the flow, the \hh\ wakes range in diameter  from 2\arcsec\  
to 8\arcsec\  ($7 \times 10^{14}$ to $3 \times 10^{15}$ cm)  with limb-brightened rims 
less than 1\arcsec\  ($< 3 \times 10^{14}$ cm) wide (Figure \ref{fig4}).   
The half dozen major  finger clusters in the northwest are up to 60\arcsec\   ($\sim$ 0.1 pc) long.  
The wakes in the inner part of the flow are narrower, tend to be shorter, and are 
more numerous resulting is a high degree of overlap along the line-of-sight.   
The \hh\ emission tends to be fainter or disappears near the wake-tips where it is 
replaced by \Feii\ emission.    

Most of the \Feii\ emission in the BN/KL outflow originates from the fingertips which 
are faint or invisible in \hh .    Dozens of  fingertips are visible  in the 1.64 \um\ \Feii\ line.   
The two blue vectors in Figure \ref{fig3} point to the two  brightest \Feii\ features which are also
bright in visual wavelength shock-excited emission lines such as \oi\ and \sii\  and are
designated as HH~201 and 210.    These fingertips are very faint in \hh .   
Figure \ref{fig4} shows a color version of  the `\hh\ fingers'  region in 2013 from GSAOI 
showing \hh\ (red),  \Feii\  (green), and the broadband light in K$_s$ (blue). 

\subsection{Proper Motions of Selected Knots}

Previous analyses of  multi-epoch ground-based images have shown that the highest 
proper  motion knots in  the BN/KL outflow are most distant from OMC1 \citep{JonesWalker1985,LeeBurton2000,Bally2011}.  
Figure \ref{fig5} shows a difference image obtained  by subtracting the 1999 epoch 
Subaru \hh\  image of \citet{Kaifu2000} from our 2013  GSAOI \hh\   image.    Although the
registration of the two images is not perfect due to small-scale distortions in the Subaru image
(this image was assembled from a mosaic of smaller images), the overall pattern
of expansion away from the OMC1 core is obvious.

Comparison of the 2013 epoch images in the sub-field containing the brightest part of the
PA $\sim$ 340\arcdeg\ to 350\arcdeg\  finger with 2007, 2008, 2009 Gemini North 
NIRI images show large proper motions of  the \Feii -dominated fingertips pointing
away from the OMC1 core.   This field contains  several compact, high velocity  \hh\  
knots with sub-arcsecond to arcsecond diameters and proper motions similar to
the \Feii\  fingertips.      Additionally, the \hh\ wakes are expanding at right angles
to a line connecting the fingertips to OMC1.    The fastest transverse proper motions 
are found close to the fingertips and decline along the parts of the wake closer to OMC1.

The most reliable proper motions were measured in this field using
the AO-assisted images from  2013 and 2007 that are separated by 2,125 days.  
Figure \ref{fig4} shows a color composite image of this field.  
Figures \ref{fig6} and  \ref{fig7}  show  difference images in   \hh\  and \Feii\  
formed by subtracting a de-distorted, intensity matched and registered
2013 GSAOI images from the corresponding 2007 NIRI images.  
Figures \ref{fig8} and \ref{fig9}  show closeup views of the central part of the
field shown in  Figures \ref{fig6} and \ref{fig7} that contains a bright bow shock. 
These  images show 200 to 300 \kms\ proper motions of the fingertips
in both \hh\ and \Feii .  
 
Figure \ref{fig8}   shows  that in addition to the large proper motions of the 
fingertips,  the \hh\ wakes are spreading with velocities 
ranging from less than 20 to about 80 \kms\  in a direction orthogonal to the motions 
of the  fingertips.     The two regions marked W1 and W2 show
the locations of slow and fast transverse spreading, respectively. 
The amount of transverse motion between the two epochs at these two 
locations are indicated by the  closely spaced dashed lines. 
The spreading velocity decreases with increasing distance from the fingertips. 
Figure \ref{fig9}  demonstrates that these spreading \hh\ regions are located 
behind a  fast-moving \Feii\ bullet whose locations in 2007 and 2013 are indicated.

The closeup \hh\ images of this region (Figures \ref{fig6} and \ref{fig8}) 
show a  high-velocity compact clump (HVCC) which exhibits a proper 
motion of $V_{HVCC} = 300 \pm 20$ \kms\   over the 6 year ($1.85 \times 10^8$ second)  
interval between the acquisition of the Gemini North and South images in 2007 and 2013.   
This HVCC  is located at projected  distance of 100\arcsec\ from the current location of
source I and d$_{ejection}$ = 97\arcsec\  (0.195 pc) from  the suspected location of sources 
I and BN more than 500 years ago  prior to their dynamic ejection.   Assuming no 
deceleration, the dynamic age of this knot is  
$t_{dyn} = d_{ejection}/ V_{HVCC}  \approx  640 \pm 30$ years (see Table~2).    
Comparison of HST images in  \oi\  and \sii\  taken on MJD 50170 show that the 
HVCC is located about 5\arcsec\  south of HH~207 which is also bright in our
\Feii\ images.   In  \Feii ,    HH~207 has a proper motion  of  229 \kms\ (Table~2).   
Measured on Hubble Space Telescope images in the visual wavelength range, HH~207
had proper motions  of   167 \kms\ in \sii ,  203 \kms\ in H$\alpha$, and 291 \kms\ in \nii 
\citep{Doi2002}.     Using the velocity difference between the \hh\ and \Feii\ proper 
motions of the HVCC and the \Feii\ emission associated with HH~207 ($\sim$68 \kms )
and the current projected separation of  5\arcsec\  indicates that the HVCC may
catch up to the moving \Feii\  emission region in about 140 years (assuming that they
are not displaced along the line of sight). 

Figure \ref{fig8} shows that this HVCC  has undergone considerable transverse spreading 
over the past 6 years.    The HVCC is either  subject to photometric variations or  
experiencing significant ``sideways splashing"  of post-shock material.    
The morphological evolution of the HVCC suggests that it may have been
decelerated significantly from its ejection velocity.   
Several less-prominent HVCCs exhibit motions of 200 - 300 \kms . 
Measured motions of a sample (subset) of fingertips, \hh\  wakes, and compact, fast
knots are given  in Table 2.    

While the fingertips with 100 to 300 \kms\  motion (best seen in \Feii )  are moving 
along a direction which points directly away  from the OMC1 core,  the \hh\ wakes 
behind each \Feii\  feature are spreading at right angles to the fingertip proper motions.  
The wake spreading-velocities tend to decrease with increasing distance behind 
each  fingertip.    Within a few arcseconds of  the fingertips,  some \hh\ wakes show 
transverse expansion as fast as 80 \kms\   orthogonal to the wake orientation and proper 
motion of the fingertip.   Spreading speeds decline to under 20 \kms\  (the measurement
limit) at distances of more than 10\arcsec\ to 15\arcsec\ behind the fingertip. 

The transverse spread is consistent with a simple  model of a disturbance created by the  
passage of a high density  compact object through the medium as shown below 
by numerical simulations.  
Let $y$ be the distance from the tip of a finger to the location where the wake half-width
(width of the wake divided by 2)  and transverse expansion speed are measured.  
The ratio of the transverse spreading-velocity, $V_{perp}(y)$ (one-half  of the velocity 
with which the two sides of a wake are moving apart) to  the fingertip 
proper motion, $V_{PM}$, is  comparable to the half-width of the wake at a given location,  
$X_{perp}(y) $,  divided by the distance of that location from the fingertip, $Y_{PM}(y)$.  
Thus, $V_{perp} (y) / V_{PM} \approx  X_{perp}(y) / Y_{PM}(y)$.     Both the fingertip
and spreading velocities are highly supersonic. 

As shown by previous ground-based measurements, the largest proper motions are seen
at the largest projected distance from the OMC1 core and the suspected locations of
the BN object and radio source I about 500 years ago.   The northwestern fingertips   
located at the greatest distance from the OMC1 core exhibit the largest 
proper motions (200 to over 300 \kms ) and tend to show a pattern of
increasing maximum  velocity with increasing distance from the core.     However 
the region also contains a number of slower-moving features with proper motions
ranging from under 100 to about 200 \kms .    It is likely that these features have
experienced significant deceleration as they interact with the surrounding medium. 
The fastest knots and fingertips have dynamic ages consistent with ejection between 
450 and 600 years ago.

\citet{Doi2002}  measured visual
wavelength proper motions for Herbig-Haro (HH) objects protruding into the 
photon-dominated region (PDR) behind the Orion Nebula.    Two of the brightest \Feii\ features
correspond to HH~201 northwest of OMC1 and 210 north of OMC1.    \citet{Doi2002}
measured proper motions of 312 to 315 \kms\ for various components of HH~201 and
and 309 to 425 \kms\ for knots in HH~210.     \citet{Grosso2006} detected X-rays from the 
wake of  HH~210, which is the highest proper motion finger  in the OMC1 outflow and one 
of the relatively few visible at visual 
wavelengths, thereby demonstrating that at least some of the fingers contain hot,  X-ray 
emitting plasma.   Additional HH objects are associated with the Orion fingers including
HH 205 to 209, and HH 601 to 607  with proper motion velocities ranging  from  100 to 
300 \kms\  \citep{Doi2002}.  All of these features have large negative radial velocities.
For example, HH~201 has \Vlsr\ $\sim$ $-$260 to $-$284 \kms  \citep{Doi2004}.

\subsection{Constraints on the ejection mechanism}

A number of models have been proposed to explain the \Feii\  fingers and \hh\ 
wakes in the  BN/KL outflow.     The fingertips contain  \Feii\ knots and bow shocks  
with diameters of  1\arcsec\ ($\sim400$ au) or less,  estimated by measuring the 
separation  between points  where tangent lines on the leading edges of the shock 
on either side of the tip are at an angle of 90\arcdeg\  with respect to each other.    
This dimension is used as an estimator for the size-scale of the material powering 
each shock which may be the working surface of a jet \citep{SmithRosen2007},  
a wind which has experienced instabilities \citep{StoneXuMundy1995}, 
or a dense bullet \citep{AllenBurton93}.    The tangent lines become increasingly 
parallel (greater than  90\arcdeg ) in the wake. Most of these \Feii\ bullets and 
the \hh\ HVCCs  are located more than  100\arcsec\ from their ejection site.    

The individual fingers resemble the shocks  produced by accretion-powered jets or 
collimated flows powered by young stellar objects as they sweep up the  ambient  
medium \citep{ReipurthBally2001}.   However the presence of many dozens of
individual fingers and over 100 shocks with similar dynamic ages moving
away from the OMC1 core in all directions suggests that a single, brief event 
is more likely.   
 
The multitude of fingers and wakes  could be powered by a  fast  wide-angle 
wind which experienced instabilities  and  broke into a multitude of  protrusions 
\citep{StoneXuMundy1995,McCaughrean_MacLow1997}.    Rayleigh-Taylor (RT)
instabilities can produce fingers of fast ejecta surrounded by slower clumps of dense 
gas if the wind velocity increases with time on a time-scale shorter than the 
crossing time of the wind in the wind-dominated region.     Alternatively,   
a steady wind running into into a stationary  medium with a density profile  
decreasing faster than $r^{-2}$ can be subject to RT instabilities.   Neither an 
accelerating wind, nor a wind running into a rapidly decreasing  density gradient 
is  expected to produce the approximately Hubble-flow type  behavior with the  
maximum velocity of the ejecta  increasing linearly with increasing 
projected  distance from the source.    

The  fingers and wakes may be driven by high-density, compact ejecta
(`bullets'  or HVCCs)  as originally suggested by \citet{AllenBurton93}.    
Similar bullets have been detected in other astrophysical contexts such
as supernova remnants where hydrodynamic and cooling instabilities fragment
an expanding stellar envelope \citep{Fesen2011,MilisavljevicFesen2013}.  

Assuming the ejection occurred $\sim500$ years ago at approximately the
location of the BN/KL infrared nebula 
\citep{BallyZinnecker2005,Zapata2009,Bally2011,Goddi2011},
the most distant knots to the northwest provide the most stringent 
constraints on the ejecta properties.    These knots have traveled
$\sim0.28$ pc in projection in $\sim$500 years.  Thus,  the time-averaged 
velocities must be greater than  $\sim$550 \kms.   For an explosive origin, 
the fingertip proper motions should increase linearly with increasing distance 
from the launch region.    While a few \hh\ features have proper motions between 300
and 350 \kms ,  most show lower velocities, especially closer to the OMC1 core.   
Because the faster motions are at least 20\% slower than expected for a 500 
year-old explosion,   the ejecta  have either decelerated or were ejected earlier.   
The absence of faster ejecta in \Feii\  and \hh\  could be a consequence of deceleration,
an older time of ejection, or a selection effect resulting from excitation 
conditions in the shocks.       \Feii\ can be in excited into higher ionization stages and 
\hh\  may be dissociated above a critical shock speed.     Such ultra-fast  
material  should  still be revealed by their \hh\  wakes.  

If the bullets  originated  between radio sources I and BN,  they moved 
between $10^2$ to $10^3$ times their current diameters  to their present location.
The HVCC shown  in Figure \ref{fig8} had a  diameter about about 
0.2\arcsec\ (100 au) in 2007 and traversed a distance of about 100\arcsec ,  
about 500 times its own diameter,  from the suspected ejection site between 
sources I and BN.     Assuming that they are moving ballistically and not actively 
powered by  jets or winds, momentum conservation requires that such bullets must 
be $\sim$500 times  denser than the environment.    Such bullets are unlikely to 
contain sufficient mass to be bound by their self-gravity.   They may be confined 
by ram pressure, otherwise they are  likely to be expanding with a speed 
comparable to their  internal sound speeds.    

If the bullets were formed from the disruption of circumstellar disks envelopes 
within tens of au of the massive stars ejected from the OMC1 cloud core, 
their initial densities are expected to be  $n(H_2) \sim 10^{10}$ to $10^{15}$ \cmq , 
comparable to the densities of disks and inner envelopes.  Following dynamic ejection,  
bullets  are likely  to expand with a velocity comparable to their an internal sound 
speed $c_s$.  In the absence of ram-pressure confinement by the medium the bullet
is likely to sweep out a conical region given by the Mach angle,  $M \sim 2 c_s/ V_B$
where $V_B$ is the bullet velocity.   Using the observed sizes of the compact \Feii\ and \hh\ 
knots (40 to 400 au)  and a distance greater than 0.2 pc from the ejection site implies 
$c_s < 0.6$ \kms\  consistent with an H$_2$ temperature of $\sim 10^2$ K .   Although 
expansion will tend to cool the bullets below the surrounding medium,
the intense IR radiation from OMC1 and  heating by shock radiation will tend to
keep them warm.   It is likely that many bullets are moving into regions already 
shocked, heated, and partially evacuated by faster, previously ejected material.  
Such clumps will  interact with a post-shock medium having a lower-density than 
the ambient medium, making it more likely  for the bullets to be in free-expansion
rather than  being confined by  ram pressure.    

The  properties of the \hh\  wakes  constrain the density of the 
medium into which the suspected HVCCs are moving.   The wakes have widths that are 
an order-of-magnitude wider than the \hh\ HVCCs and \Feii\  fingertips (2\arcsec\ to 10\arcsec\ 
with most being near the lower-end of this range).   Their formation requires 
that the heated layer behind the forward shock moving into the ambient medium
has a cooling length  larger than the clump  (or the width of
the  jet beam) \citep{Blondin1990}.       Figure \ref{fig10_cartoon}  shows a cartoon of a 
dense bullet moving through a medium.    A forward shock-heated  ambient medium  
streams around the moving clump and drives  lower-velocity shocks sideways 
into the ambient medium.   A much slower, reverse shock propagates into the clump
and compresses its leading edge. 

The post-shock temperature immediately behind a  shock is  given by 
$T_{ps} = 3 \mu V_s^2/ 16 k$ where $V_s$  is the 
shock speed.   \citet{Blondin1990} give the cooling distance as
$d_{cool} = V_s t_{cool} / 4$ $= 9 \mu V_s^3 / 64 n_0 \Lambda (T_{ps})$
$ \approx 4.5 \times 10^{16} V_{100}^{4.0} / n_0$ 
where $\mu$ is the mean molecular weight of the pre-shocked gas particles, 
$\Lambda (T_{ps})$ is the cooling function,  $V_{100}$ is the
shock velocity in units of 100 \kms , and $n_0$ is the pre-shock particle density.   
More recent numerical studies of the cooling function give 
$d_{cool} \approx 5.5 \times 10^{17} V_{100}^{4.4} / n_0$  for $80$ \kms\
$< V_s <$ 1,200 \kms\ \citep{Draine2011}.  Thus, a $V_s$ = 300 \kms\ shock
moving into a density $n_0 = 10^4$ \cmq\ has a cooling length 
$L_{cool} \sim 7 \times 10^{15}$ cm (470 au).  Thus, for ambient medium 
densities between $10^3$ \cmq\ and $10^4$ \cmq\  typical for the 
`Integral Shaped Filament' in the Orion A molecular cloud located behind the 
Orion Nebula, these cooling  lengths correspond to 1.1\arcsec\   to 11\arcsec\   
at the distance of Orion, and are larger than the HVCCs and \Feii\  knots.   

For a dense HVCC driving a forward shock into the medium with  speeds of 150 to 500 
\kms , the hot ($\sim$1 to 10 MK) plasma will splash sideways to produce a wide
bow-shaped wake.    In the OMC1 rest frame the 30 to 300 year cooling time corresponds 
to a wake-length  $L_{wake} \sim 3 \times 10^{16}$ to $3 \times 10^{17}$ cm, the
latter being comparable to the lengths of fingers in the north and west parts of the 
OMC1 outflow.    The sideway expansion of the hot plasma into the surrounding 
medium can then drive  a slower ($V_{side} < $ 100 \kms ) side shock.   The  \hh\ emission
is likely to be produced in the swept-up, compressed, and shock heated layer  
behind the shock.   However, intense UV radiation produced by the shock hot plasma
may also excite \hh\ molecules outside the wake, resulting in fluorescent 
excitation.

The reverse shock propagating back into a dense bullet will be much weaker than the 
forward shock owing to the bullet's higher density.  For a forward shock moving into a medium 
with density $n_0$ with velocity $V_S$, the speed of the reverse shock moving into the 
bullet having a density $n_B$ (in the frame of the bullet) is  $V_B \sim V_S (n_o/n_B)^{1/2}$.  
For a density contrast ratio  $n_B / n_o \approx 10^3$,  the reverse shock 
propagating back into the bullet will have  speed $V_B \sim 0.03 V_S = 10$ \kms\ 
for a forward shock speed of 300 \kms .

\subsection{Numerical modeling}

We use the Eulerian adaptive mesh refinement (AMR) hydrodynamics + N-body 
code,  \texttt{Enzo} \citep{Bryan2014} to model the propagation and evolution 
of bullets and the shocks they produce as they migrate through the ambient.  
The simulation set up is analogous to that of  \citet{Silvia2010, Silvia2012}  in 
which an over-dense, spherical cloud is embedded in a uniform ambient medium.  
For the simulation presented here, we have assumed an adiabatic equation of 
state that ignores any effects radiative cooling may have in this scenario.  
More detailed calculations using a realistic cooling curve for the post-shock 
gas will be presented in a future paper.

In order to track the largest possible head-tail structure as the cloud is disrupted, 
without losing information due to flows that might exit the computational domain, 
we give the ambient medium a uniform velocity such that it washes over the cloud, 
rather than moving the cloud itself.  As the medium flows past the cloud, it drives 
shocks into the cloud interior and ablates material from the cloud-medium 
interface -- producing a tail of material that is carried downstream.  In order to 
capture the evolution of the cloud over a large physical volume while minimizing 
the computational cost of the simulation, we use a coarse-resolution root grid with 
a physical cell size of $\sim$78~au per cell and eight levels of adaptive mesh 
refinement such that the cloud structure is resolved at a peak resolution of 
$\sim$0.3~au per cell.  The full computational volume is 7500 
$\times$ 2500 $\times$ 2500~au.

Figure \ref{fig11} shows a comparison between one of the H$_2$ fingertips and the 
numerical model showing the density structure of a dense, supersonic cloud 
that  has moved approximately 500 
times its initial diameter through a uniform medium.   At the start of the simulation, 
the medium is initialized with a number density of $n = 10^3$~cm$^{-3}$, a 
temperature of $T = 20$~K, and uniform velocity of $v = 250$~km~s$^{-1}$ 
directed along the $x$-axis of the simulation domain.  The cloud is initialized 
with a density 10$^4$ times higher than that of the ambient medium, a
temperature of 100~K, and a radius of $r = 10$~au.

Figures \ref{fig12} through \ref{fig16}, made using the yt toolkit \citep{Turk2011},
show the density,  $x$,$y$,$z$-velocity 
structure,   and temperature of the bullet after 100 years of evolution. 
For this simulation, the initial velocity of the cloud with the respect to the 
medium is too slow to reach the observed positions given an ejection 500 
years ago, but is compatible with an earlier ejection time.  Additionally, 
the simulation shows significant cloud deceleration and disruption in 
just 120 years following ejection.  Together, this indicates that the bullet 
over-densities are likely to be higher than 10$^4$, or that the ISM densities 
are lower than used in the models. Prior to the dynamic ejection event, the 
forming stars probably produced accretion-powered outflows which would 
have carved cavities in the surrounding cloud through with ejecta would 
be moving. Additionally, some bullets may be propagating through cavities 
and wakes created by the leading edge of the ejected debris field.

The inclusion of cooling in the post-shock regions will tend to lower the 
transverse expansion of the shock-heated plasma to produce a narrower 
wake and a lower sideways splashing velocity. Cooling of the shock-heated 
layer produced by the slower, reverse shock moving back into the bullet will 
also tend to result in extensive fragmentation as portions of the cloud rapidly
cool and condense -- a phenomenon observed in the cloud-crushing 
simulations of \citep{Silvia2010, Silvia2012}.  Such fragments could have 
considerably longer survival times than the structures present in the 
simulation presented here.  Other physical phenomena not included in this 
numerical work (e.g. magnetic fields, thermal conduction, plasma viscosity) 
may also aid in cloud survival.  The addition of such physics is left to future work.

\subsection{Other YSOs in the field}

The line-of-sight to the OMC1 BN/KL outflow  contains both lower-mass  
protostars embedded in or near the OMC1 cloud core, and more 
mature young stellar objects in the foreground Orion Nebula.    Several dozen 
YSOs fall into the field-of-view presented here.   Several of these YSOs
drive outflows which can be seen in our images.   
 
The most prominent outflow crosses the 
northern part of the BN/KL outflow around Declination $-$05:31 at position
angle, PA $\sim 240$\arcdeg\ to $250$\arcdeg\ and indicated by a
dashed arrow  near the top of Figure \ref{fig3}.   Comparison with the  \oi\  
and \sii\   HST images presented  by \citet{Doi2002}  show that this flow 
consists of  a bow shock located at  J2000 = 5:35:06.8, $-$5:18:27. 
\citet{Doi2002} label this feature as 115-101 and 115-102 in their Figure 4 where
it is shown as part of HH~603 and measured a proper motion toward   
PA $\sim 230$\arcdeg\ with a velocity of 70 to 75 \kms .  In the \hh\ image, 
a larger  but fainter bow shock is located north of 115-101 and 115-102 at 
J2000 = 5:35:11.6, $-$5:20:54.  In our \hh\ images, this 2\arcsec\  radius
bow shock shows a proper motion parallel to 115-101 and 115-102 with 
a velocity of about 70 \kms .   This feature is also faintly seen in the  
HST \oi\  and \sii\ images obtained in 1994.    Assuming that the \oi , \sii , 
and \hh\ emission arise from the same region, these images confirm the mostly 
west-southwest motion.
Finally, the \oi\ image shows a very faint bow-shaped feature at 
J2000 = 5:35:08.9, $-$5:18:10 with a morphology and orientation consistent
with the above proper motion vectors.   Apparently, this set of three west-facing
and west-moving shocks traces an outflow which originates from a yet-to-be 
identified YSO located  northeast of the OMC1 core in the Integral 
Shaped Filament of dense gas in the Orion A cloud.

The variable star  V2270 Ori (05:35:15.394, $-$05:21:14.11) located 71\arcsec\ 
north of BN along the line-of-sight to the eastern edge of our images drives a bipolar 
\Feii\ outflow at PA $\sim 34$\arcdeg\ (Figure \ref{fig17}).    The flow consists of
a pair of bow-shaped features  1.5\arcsec\ southwest and 2.1\arcsec\ northeast
of the star.   The southwest portion of the flow forms a jet  extending 
several arc seconds  beyond the southwest bow.  A curved, less-collimated 
jet-like feature extends to the northeast.   This morphology is reminiscent
of a jet produced by coverging flows at the end of a nozzle 
\citep{Canto_nozzle1980, Canto1981}.
Spectra were obtained at the Apache Point Observatory 3.5 meter telescope using
the Triplespec spectrometer (these results will be presented fully in
a forthcoming paper).  These spectra show that the southwestern jet  is
redshifted with a mean radial velocity of about
$+30$ \kms\ with respect to the rest frame of the Orion Nebula \Feii\ emission.
The northeast lobe of the V2270 \Feii\ jet is blueshifted with a radial velocity
of about $+30$ \kms .   The velocity changes 
at the position of the star, indicating that V2270 Ori is the driver of this outflow.  

There is a silhouette disk system at 5:35:14.692 $-$5:22:20.36 about 8\arcsec\  
east of  BN (Figure \ref{fig18}).  It is most evident  in the \Feii\  images, but also 
seen  in the \hh\  and Ks filters.   The arcsecond-scale bipolar nebula perpendicular 
to the shadow  likely traces reflected continuum light from the central star
embedded in and shadowed by the nearly  edge-on disk.    The shadowing 
occurs at position angle 9\arcdeg\ implying a disk axis at about 99\arcdeg .    

\section{Conclusions}

New images of the OMC1 BN/KL outflow, obtained with the Gemini South 
MCAO laser guide star adaptive optics system are presented.   The images,
obtained in the 2.12 $\mu$m \hh\ line, the 1.64 $\mu$m \Feii\ line, and in a
broad-band K$_s$ filter nearly reach the diffraction limit of the 8 meter telescope,
about 0.08\arcsec.     Complete coverage of the outflow's \hh\ wakes and \Feii\ 
fingertips reveal over 100 individual shocks in unprecedented detail.  

These images are combined with 2007 to 2009 epoch images obtained with the
Gemini North AO system, Altair, and the near-IR camera NIRI, and earlier epoch
natural seeing limited images to measure new proper motions of selected 
features and to search for morphological changes in the BN/KL outflow.

The observed morphologies of the fingers and wakes are compared with 
numerical hydrodynamic simulations of a compact, high density bullet moving through
a lower density ambient medium with a Mach number of about $10^3$.    The 
main results of this investigation are:

\begin{itemize}

\item  
Several dozen \Feii\ bright fingertips and a few sub-arcsecond diameter \hh\ 
knots show the highest proper motions.  The proper motions are 150 to
300 \kms\  at the  projected north end of the outflow complex.  

\item
The survival of \hh\ in a knot (the HVCC) moving with a proper motion or 
order 100 \kms\  indicates that it must have a density much higher than the 
surrounding medium. Over the last 6 years, this feature has shown dramatic 
morphological changes as it approaches a high velocity \Feii\ knot located  
several arcseconds father 
downstream.   The \hh\ emission from this  high-velocity compact clump (HVCC)
has spread laterally (orthogonal to its proper motion), expanding from a diameter 
less than 0.1\arcsec\ to more than 0.4\arcsec , possibly indicating the rapid
evolution of the reverse shock moving into the clump

\item
Many of the wakes show measurable expansion with a velocity of up to $\sim$
80 \kms\ orthogonal to the proper motions of the fingertips which produce them. 
The compact structure of the \Feii\ and \hh\ emission of the fingertips suggests
that they are powered by sub arcsecond to arcsecond scale bullets which have
a much greater density than the medium into which they are moving.    It is argued that
the density in the bullets must be more than three orders-of-magnitude greater than 
the background  density. 

\item
A simple model of bullet propagation is presented in which a fast forward shock
produced mega-Kelvin plasma whose cooling length is much longer than
the diameter of the bullet.  The expansion of this plasma drives shocks into
the medium at right angles to the bullet motion to produce $<$ 100 \kms\ shock
which can sweep up the ambient medium without completely destroying \hh .  These
sideways shocks driven by the expansion of the shock-heated medium are responsible
for the intense \hh\ emission in the OMC1 BN/KL outflow.   The reverse shock moving
into the bullets is slow and may excite some detectable \hh\ emission. 

\item
Preliminary numerical simulations with the \texttt{Enzo} code using an adiabatic equation 
of state reproduce the observed morphology and kinematics of the fingers. The
numerical simulations reproduce the observed sideways splash, the spreading
of the post-shock material orthogonal to the propagation of the bullets, and the 
fragmentation of the leading edge of the high-velocity debris.  Further numerical modeling
is needed to fine-tune the model parameters to better match the densities and
dynamical ages of the observed OMC1 shocks.

\textbf{Acknowledgments}:
{This research of JB and AG was in part supported by  National Science 
Foundation (NSF) grant  AST-1009847.  We thank the staff the Gemini South observatory, and 
in particular the contributions of Rigaut, F., Neichel, B., d'Orgeville, C., Carrasco, R., 
Michaud, P., and the team responsible for developing and commissioning  the Gemini
South Adaptive Optics Imager and the multi-conjugate adaptive optics system without which
these observations would have been impossible.    We thank, Richard (Deno) Stelter for 
assistance  in the initial reduction of the Gemini North NIRI images.   Many of the figures
were generated with SAOImage ds9.  Computing resources for the simulation were 
provided by the High Performance Computing Center (HPCC) at Michigan State University.
}

\end{itemize}

\bibliographystyle{apj_w_etal}
\bibliography{Orion}

\clearpage
\input{Orion_figures1}

\clearpage
\input{Orion_figures2}

\clearpage
\input{Orion_table1}

\clearpage
\input{Orion_table2}

\end{document}

%% file: Orion_figures1.tex
\begin{figure*}[htp]
\center{\includegraphics[width=1.0\textwidth,angle=0]
  {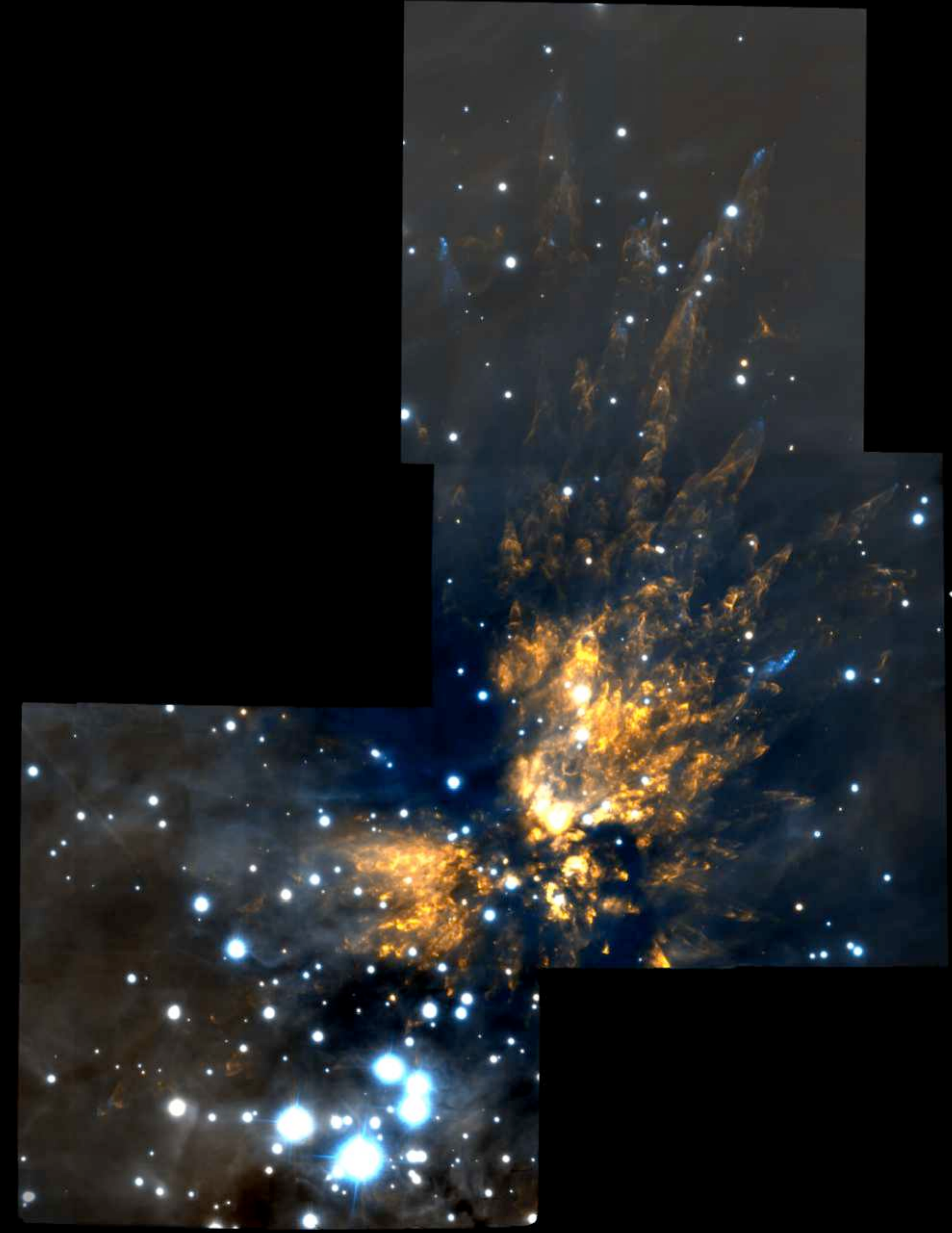}}
\caption{A wide-field image showing the OMC1 outflow in \hh\ (orange) and \Feii (cyan).}
\label{fig1}
\end{figure*}

\begin{figure*}[htp]
\center{\includegraphics[width=0.9\textwidth,angle=0]
  {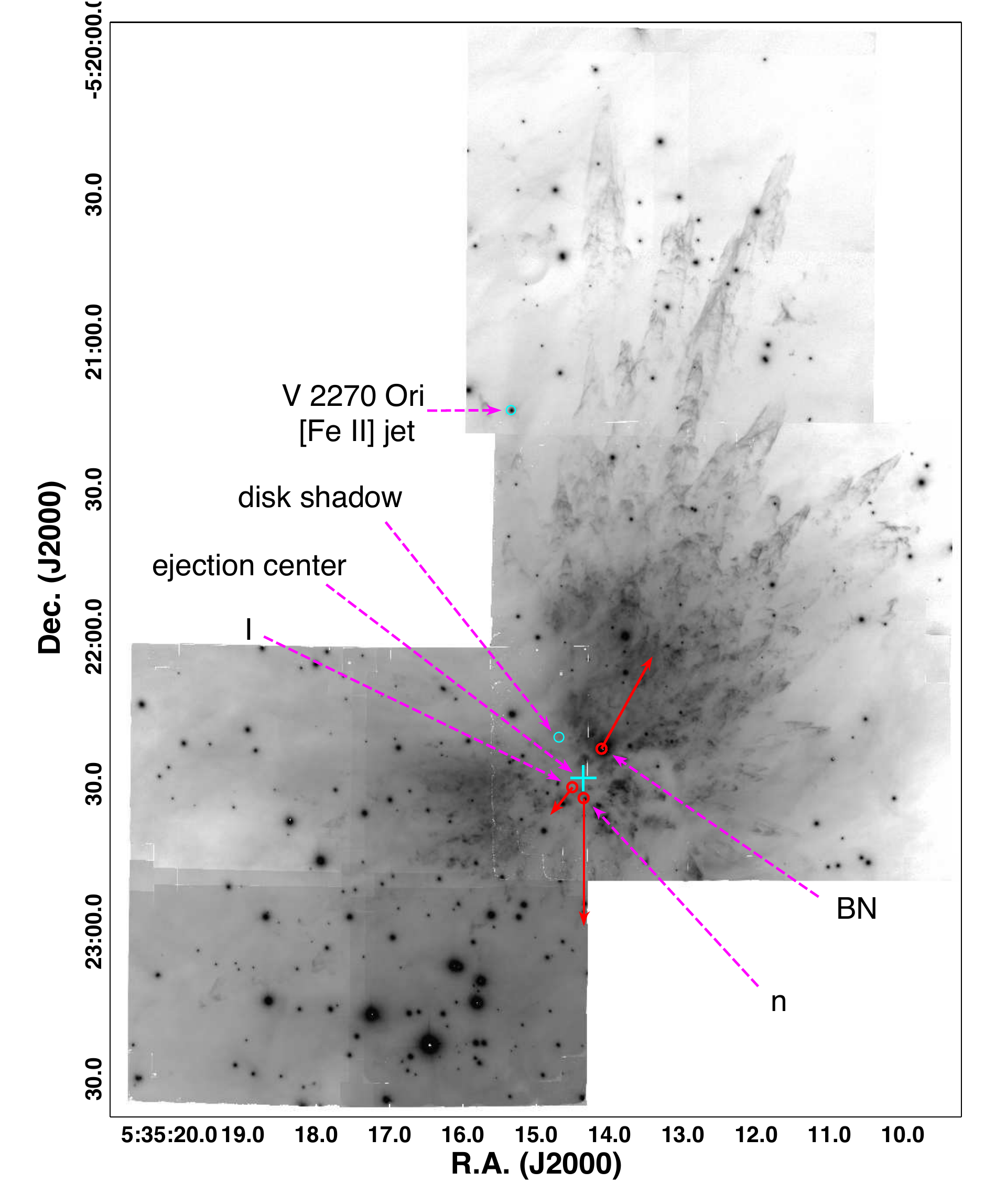}}
\caption{A wide-field image showing the OMC1 outflow in \hh\  with proper 
motions of the BN object, radio source I, and radio source n superimposed.   
The lengths of the red-solid vectors are proportional to the motions 
measured by Gomez et al. (2008) with the the lengths of the vectors in 
arcseconds equal to the motion in \kms\  (e.g. 10\arcsec\  corresponds 
to a motion of 10 \kms ).    The ejection center as determined by radio 
proper motions is shown by a cross. The locations of the disk shadow 
shown in Figure \ref{fig18} and the source of the \Feii\  jet shown 
in Figure \ref{fig17}, V2270 Ori, are indicated. }
\label{fig2}
\end{figure*}

\begin{figure*}[htp]
\center{\includegraphics[width=0.9\textwidth,angle=0]
  {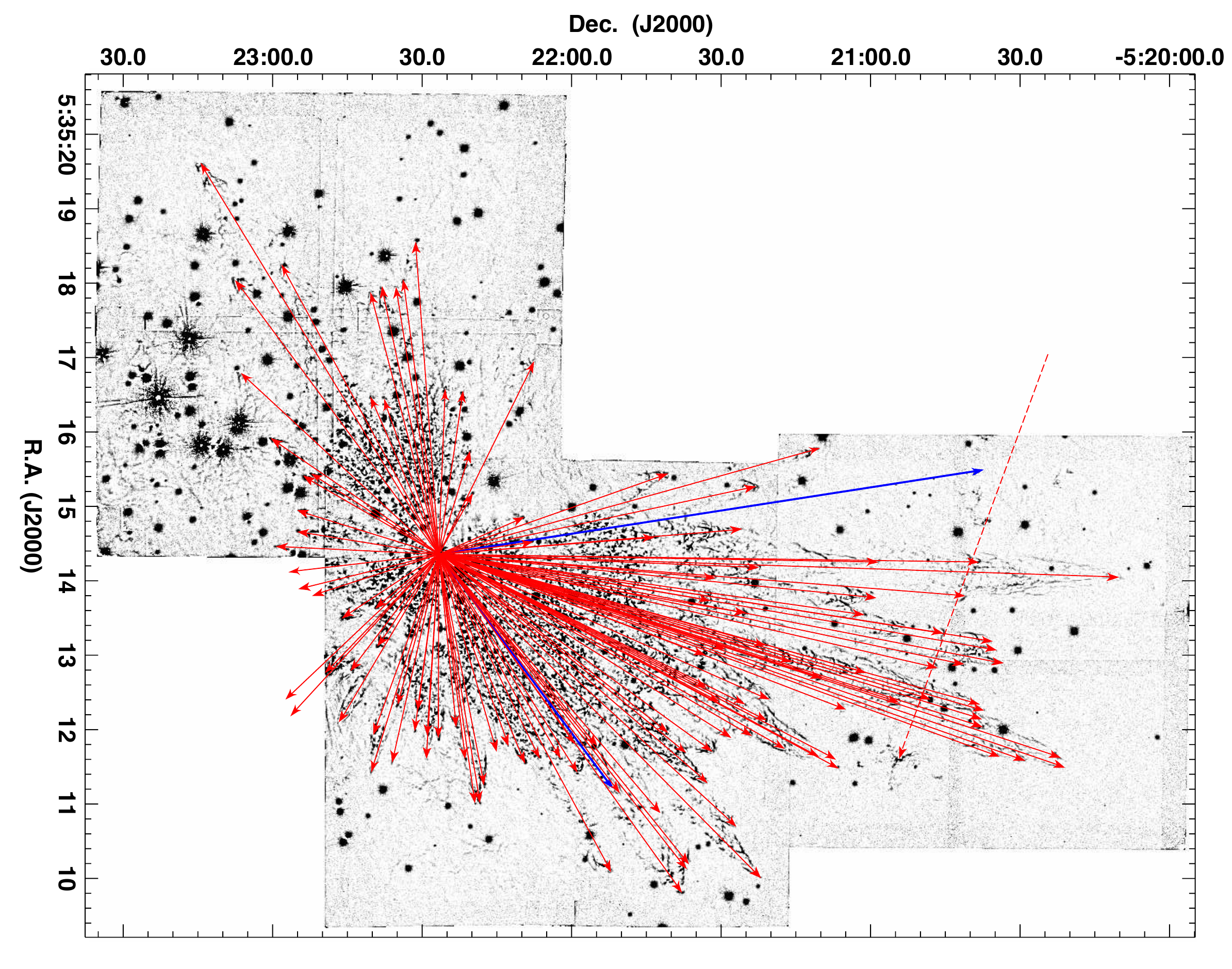}}
\caption{The outer system of fingers (vectors) in the  2.122 $\mu$m  
H$_2$  GSAOI image.  The dashed vector shows the background flow 
whose motions are towards the west.   The blue vectors mark the direction
from the explosion center to the two \Feii - dominated shocks, HH~201 and
HH~210.   The background image has 
been median filtered as described in the text. } 
\label{fig3}
\end{figure*}

\begin{figure*}[htp]
\center{\includegraphics[width=1.0\textwidth,angle=0]
  {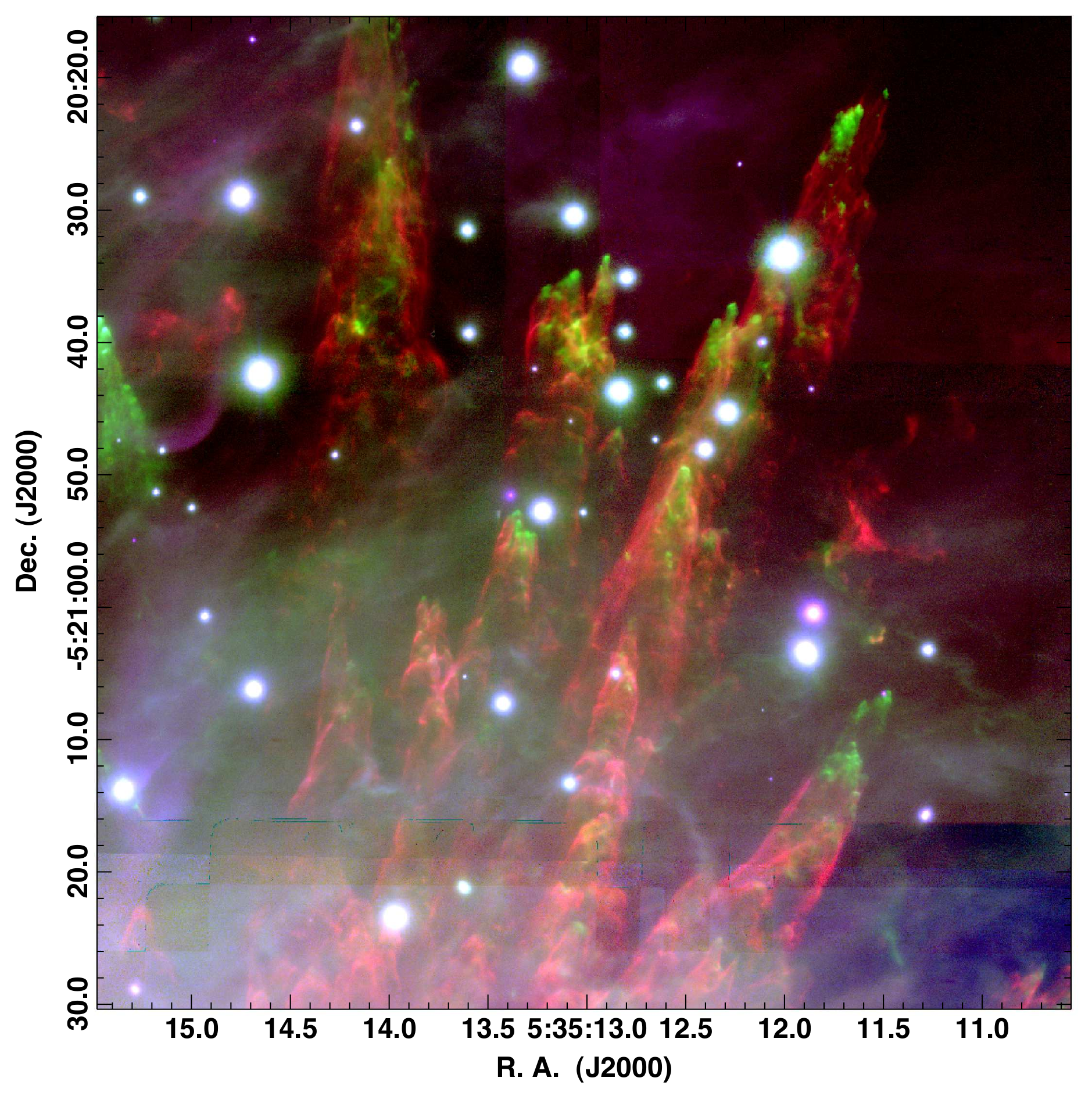}}
\caption{A color image showing the K$_s$ image (blue), the 1.644 
$\mu$m  [FeII]  image (green), and 2.122 $\mu$m H$_2$ (red)  in 
the  `H$_2$ fingers' field using the GSAOI  2013 data.   } 
\label{fig4}
\end{figure*}

\begin{figure*}[htp]
\center{\includegraphics[width=1.0\textwidth,angle=0]
  {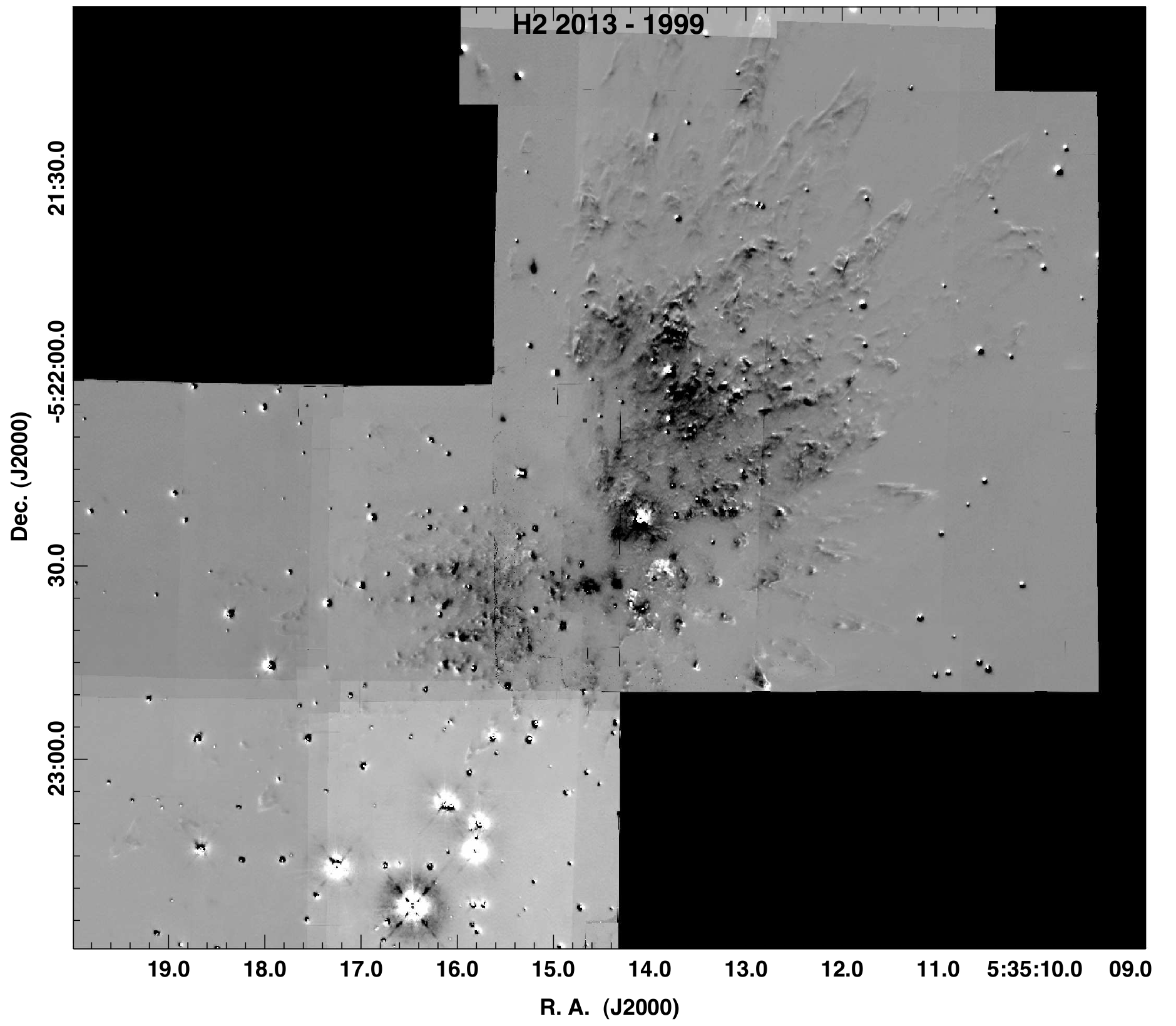}}
\caption{A 2.122 $\mu$m  H$_2$  difference image showing 
proper motions in the field imaged by Kaifu et a. (1999) with 
the Subaru telescope in 1999.   The images shows the difference 
between 2013 and 1999 epoch data.   Residual distortion 
corrections result in imperfect registration. } 
\label{fig5}
\end{figure*}

\begin{figure*}[htp]
\center{\includegraphics[width=1.0\textwidth,angle=0]
  {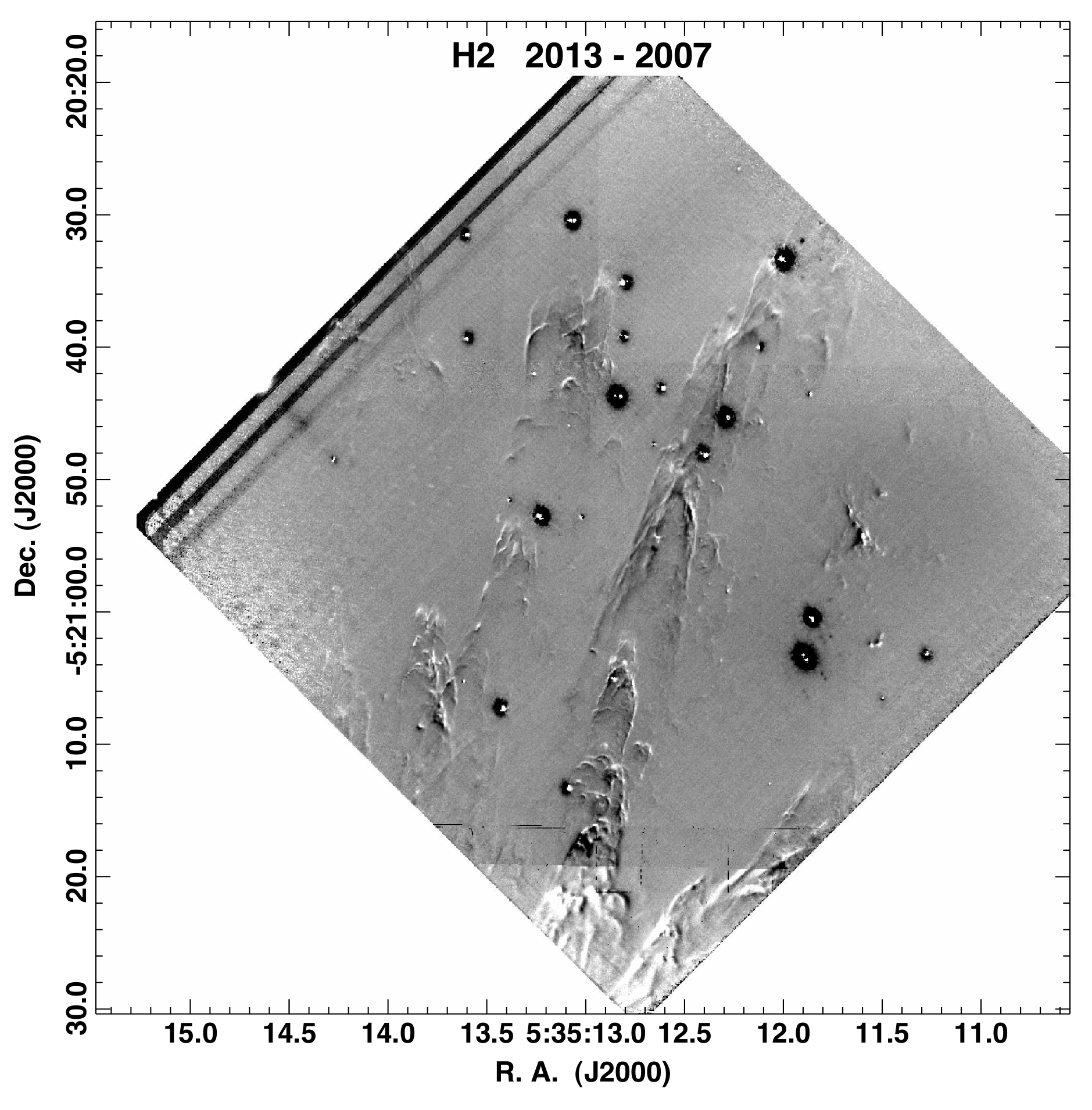}}
\caption{A 2.122 $\mu$m  H$_2$  image showing proper 
motions in the `H$_2$ fingers' field.  The image shows the 
difference between images obtained in 2013 with Gemini S 
using GSAOI and in 2007 with Gemini N using NIRI.} 
\label{fig6}
\end{figure*}

\begin{figure*}[htp]
\center{\includegraphics[width=1.0\textwidth,angle=0]
  {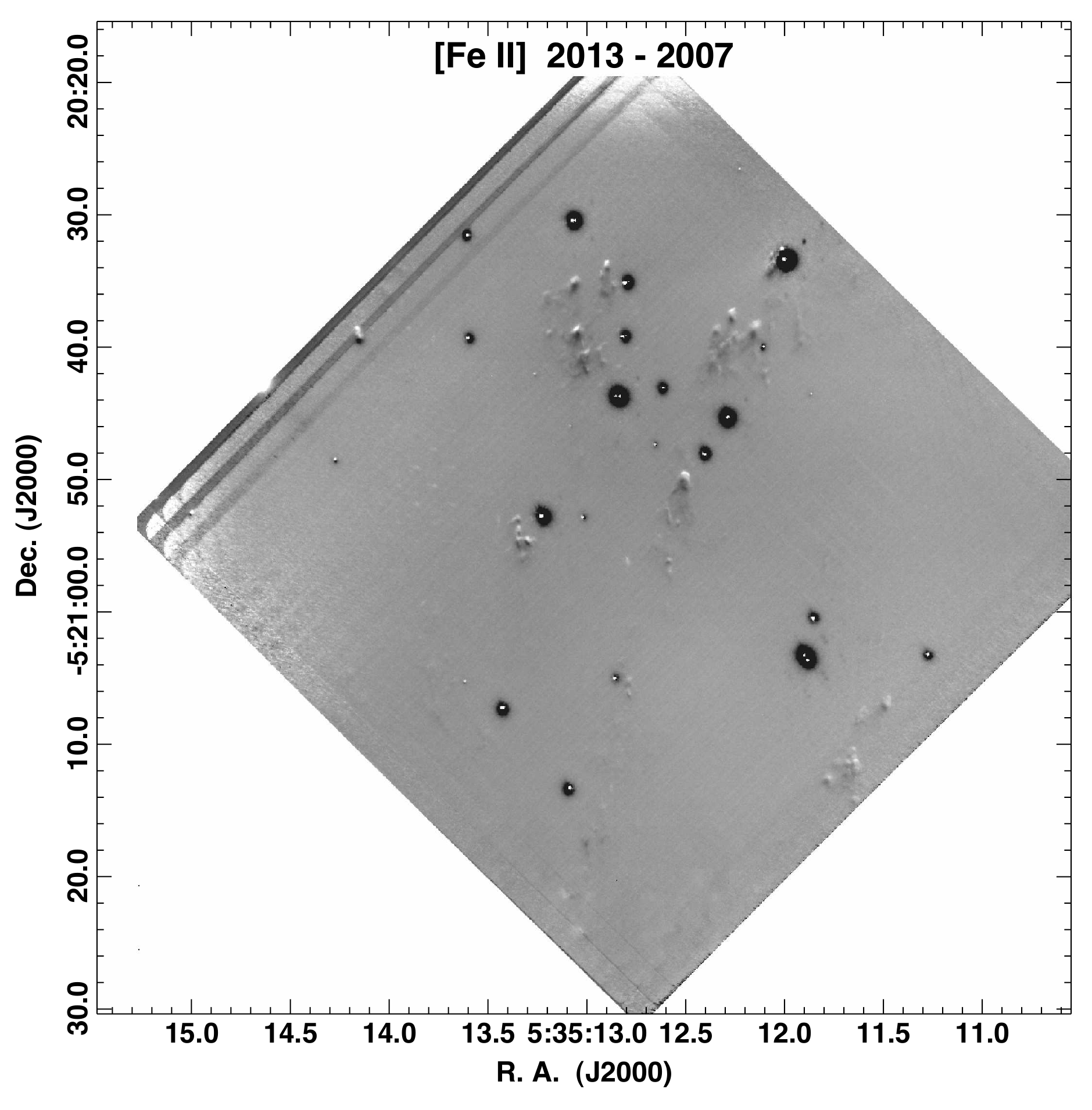}}
\caption{A 1.644 $\mu$m  [FeII]   image showing proper motions in the `H$_2$ fingers' field.  The image shows the difference between images obtained in 2013 with Gemini S using GSAOI and in 2007 with Gemini N using NIRI.} 
\label{fig7}
\end{figure*}

\begin{figure*}[htp]
\center{\includegraphics[width=1.0\textwidth,angle=0]
  {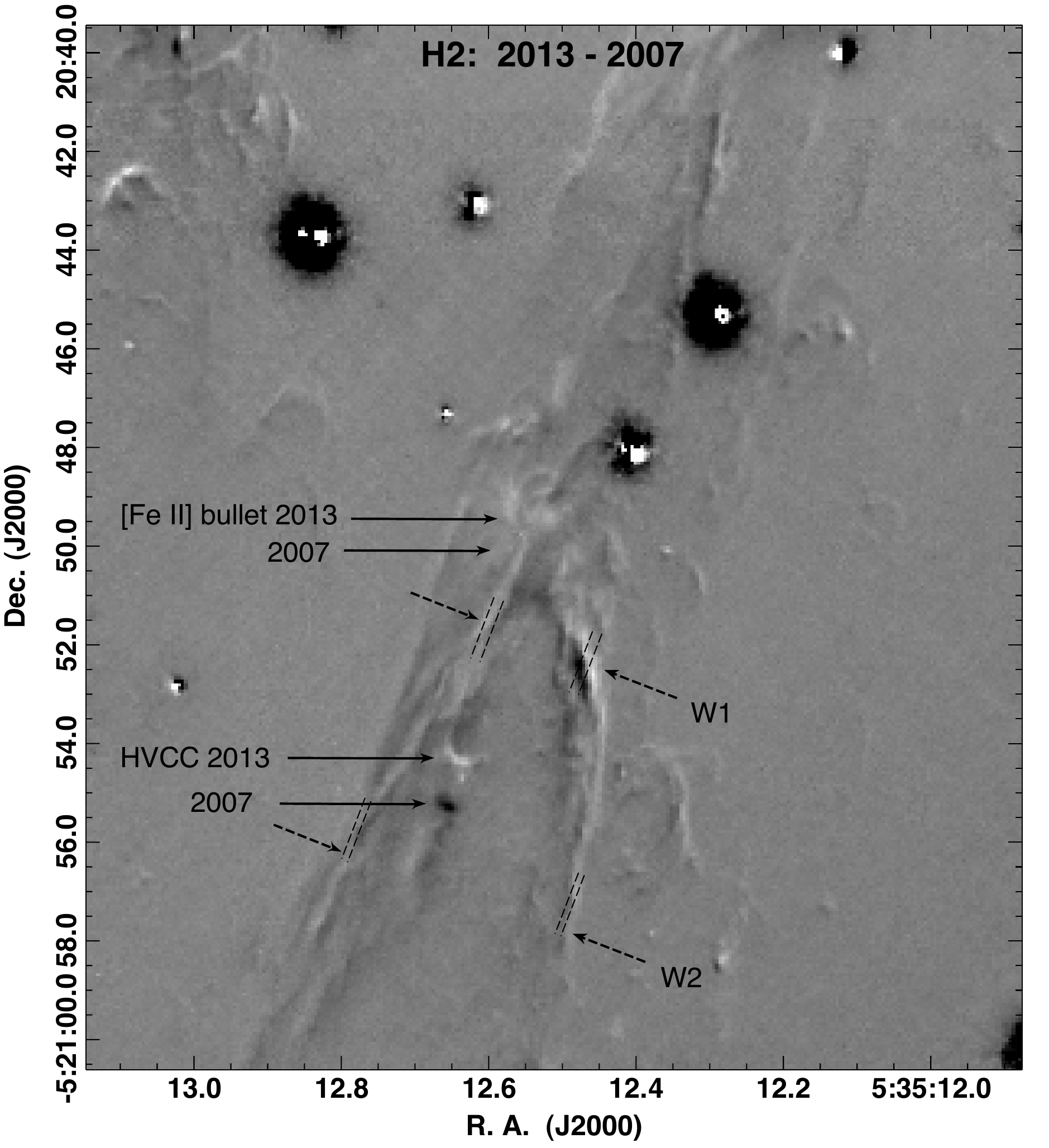}}
\caption{A closeup view of  the  main H$_2$ finger  and the HVCC in the `H$_2$ fingers' field showing  the difference between images obtained in 2013 with Gemini S using GSAOI and in 2007 with Gemini N using NIRI.} 
\label{fig8}
\end{figure*}

\begin{figure*}[htp]
\center{\includegraphics[width=1.0\textwidth,angle=0]
  {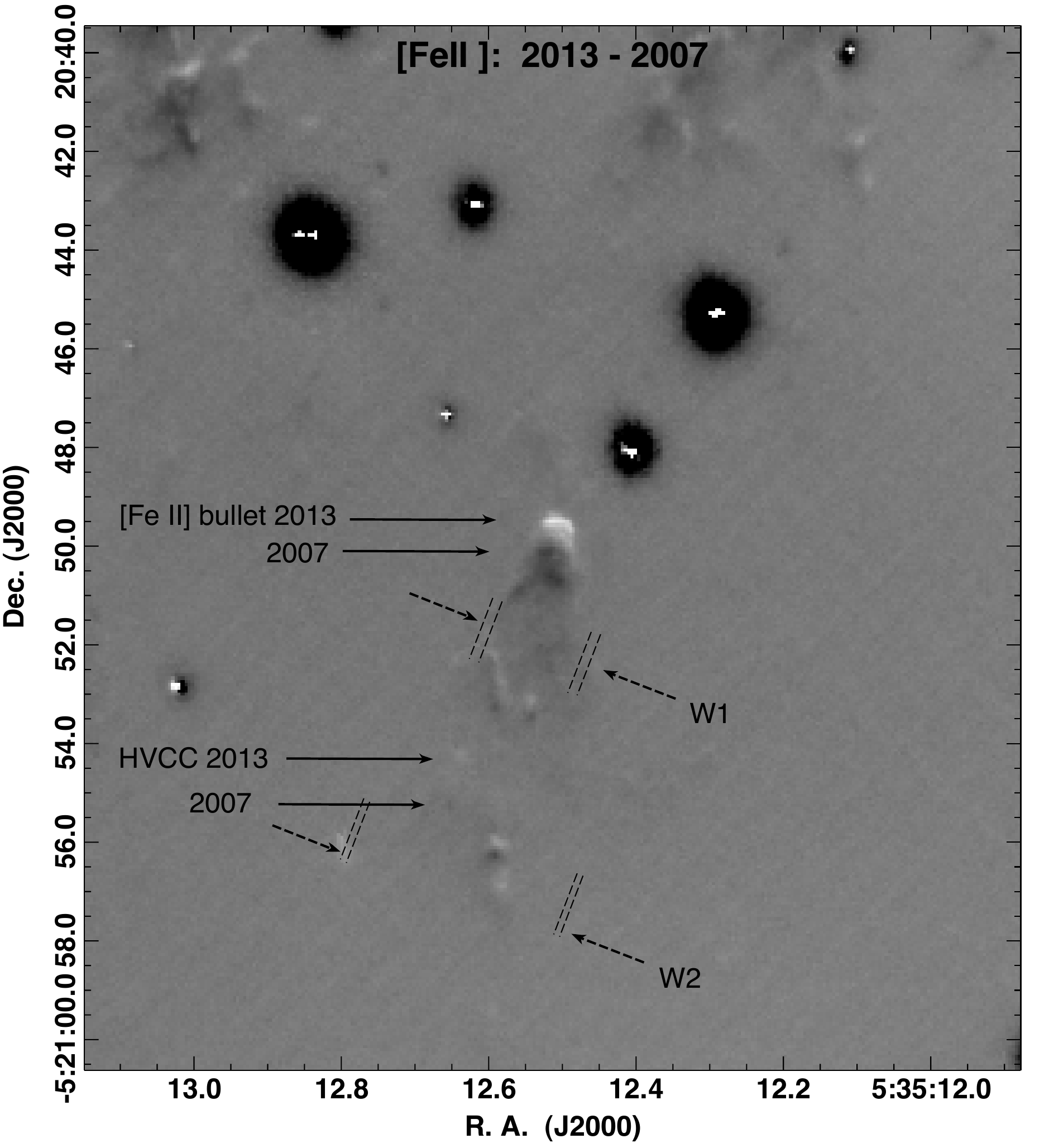}}
\caption{Same as Figure \ref{fig7} but for \Feii .} 
\label{fig9}
\end{figure*}

\begin{figure*}[htp]
\center{\includegraphics[width=0.8\textwidth,angle=0]
  {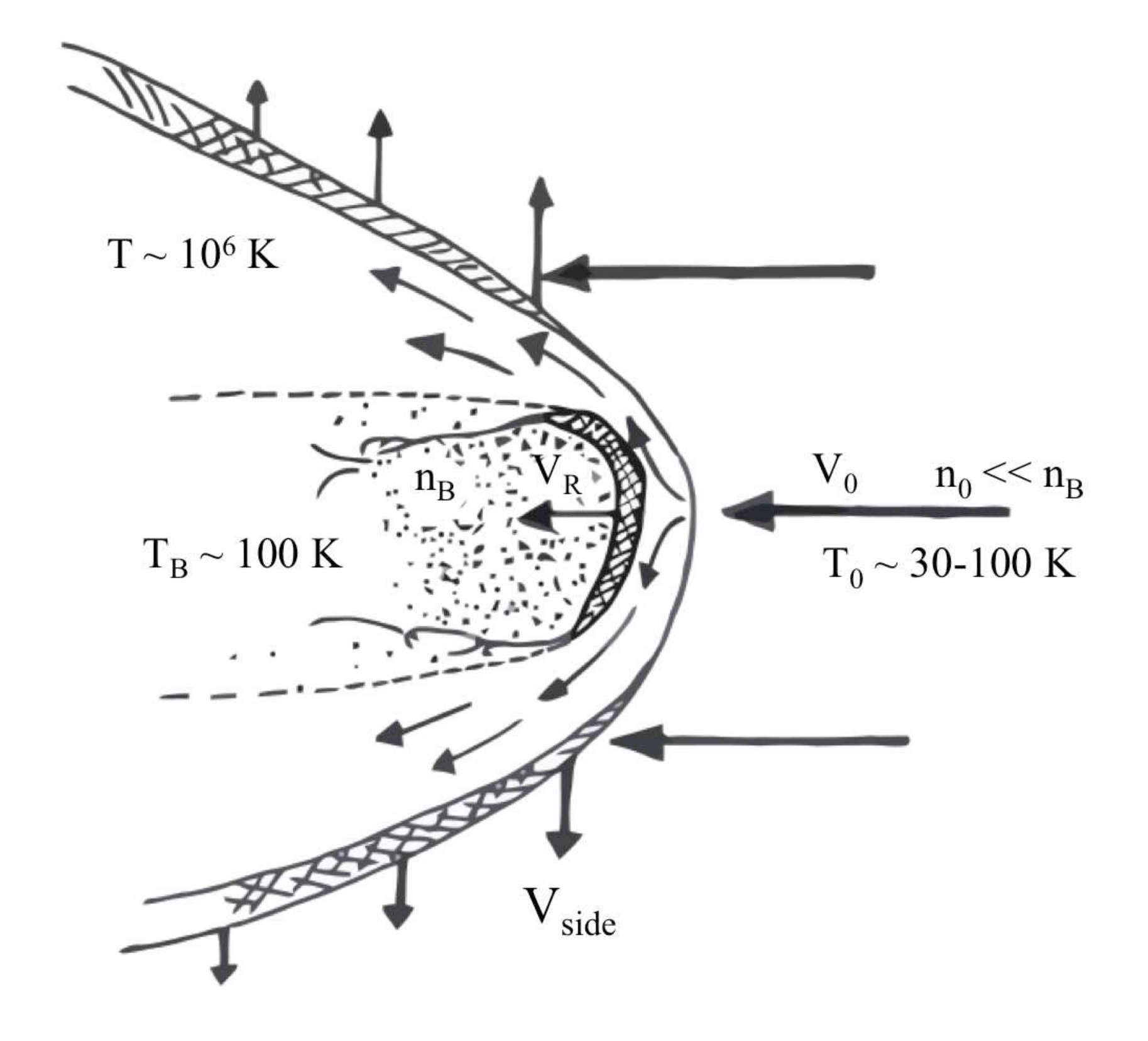}}
\caption{A cartoon showing the shock structure formed
by a hypersonic, dense bullet moving through a cold medium. The flow
vectors are shown in the rest-frame of the bullet.    The forward shock velocity $V_0$
is nearly identical to the bullet velocity and heats the post-shock ambient
medium to at temperature $\sim10^6$ K.    The reverse shock propagated back into
the bullet with a speed $V_R \sim V_0 (n_o / n_B)^{1/2}$, where $n_0$ is the density
of the ambient medium and $n_B$ is the bullet density.    The transverse expansion
of the hot, high-pressure, post-shock medium sweeps-up the colder, lower-pressure
ambient medium to form the \hh\ wake.
Only the transverse component of the vector $V_{side}$ is shown.  The fluid velocity 
of the material swept-up by the expansion of the hot, shocked ambient medium
is the vector sum of the pre-shock ambient medium (as seen in the reference 
frame of the bullet), and the velocity of the sideways splash produced  by the
expansing hot plasma.
} 
\label{fig10_cartoon}
\end{figure*}

\begin{figure*}[htp]
\center{\includegraphics[width=1.0\textwidth,angle=0]
  {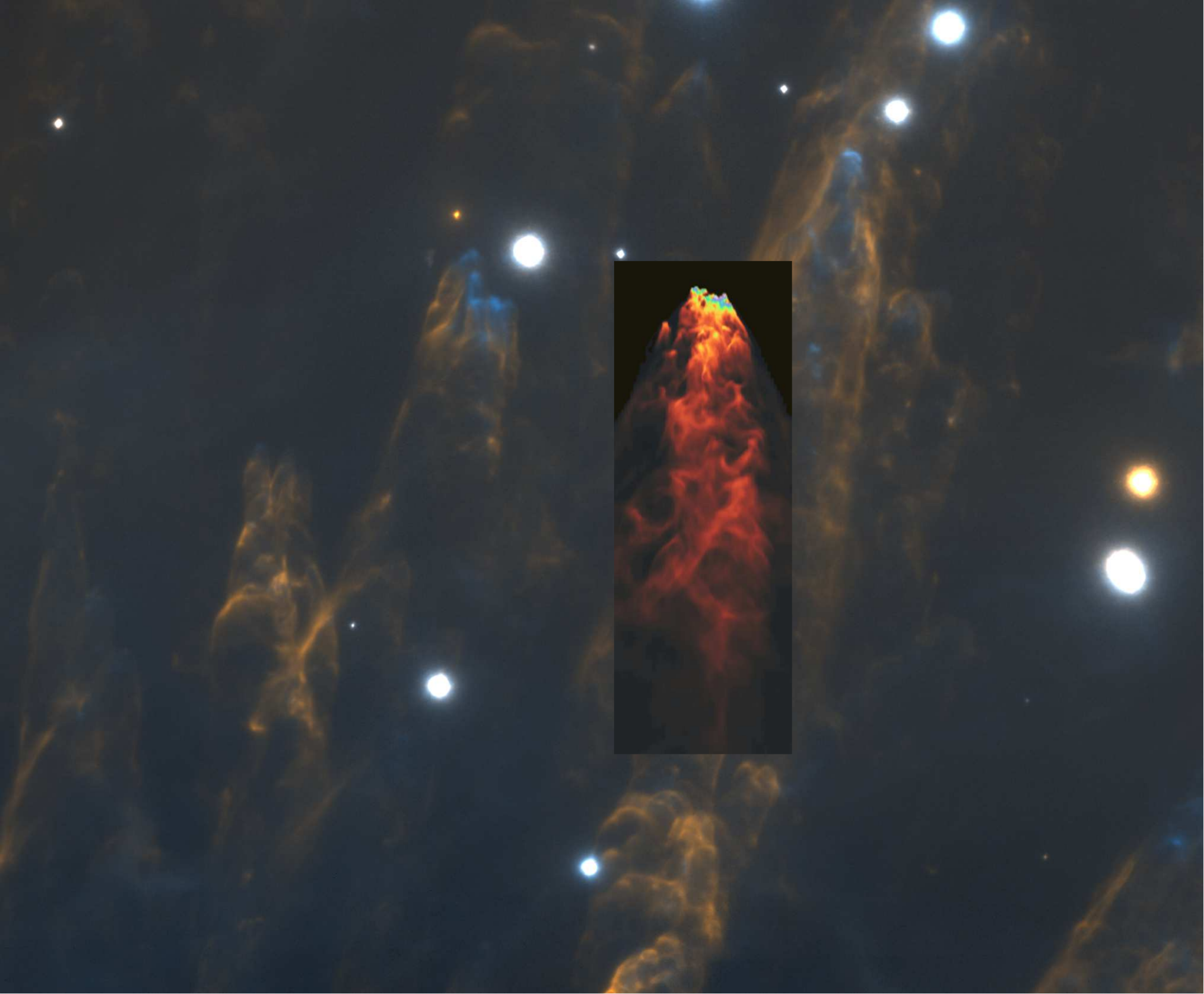}}
\caption{A numerical simulation of a spherical object with a density $10^3$ times
higher than the background density, moving with a velocity $10^3$ times faster than
the sound speed in the medium.  The frame shows the projected density
distribution after the objects has moved about $10^3$ times its initial
diameter.  The simulation projection is superimposed next to one of the fingers
shown in Figure 1.} 
\label{fig11}
\end{figure*}

\begin{figure*}[htp]
\center{\includegraphics[width=1.0\textwidth,angle=0]
  {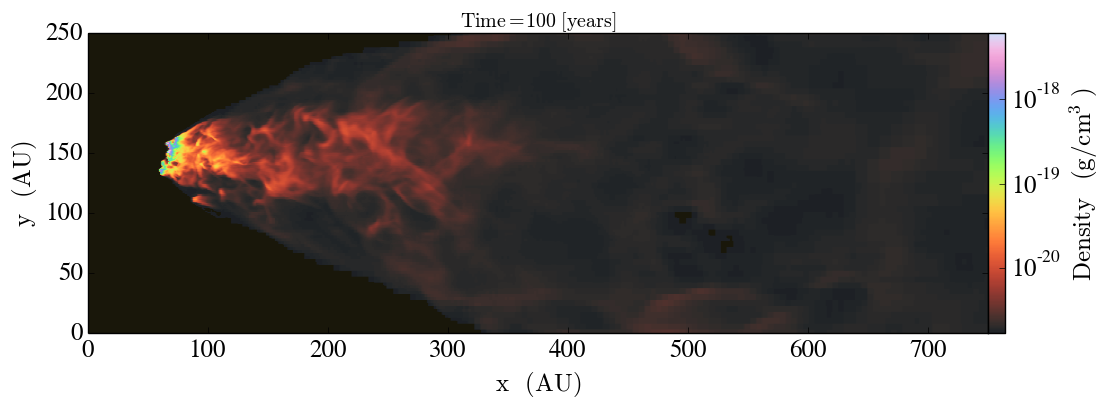}}
\caption{A density-weighted projection of gas density from the Enzo simulation
described in Section 3.3 at t = 100 years. Projected values are
computed such that 
$\rho_{proj} = \sum \limits_{i} \rho^2_{i} $ /  $\sum \limits_{i} \rho_{i}$ where 
$\rho_{i}$ is the density,
$i$ denotes the cell number, and $\rho $  is the density in cell $i$.  
} 
\label{fig12}
\end{figure*}

\begin{figure*}[htp]
\center{\includegraphics[width=1.0\textwidth,angle=0]
  {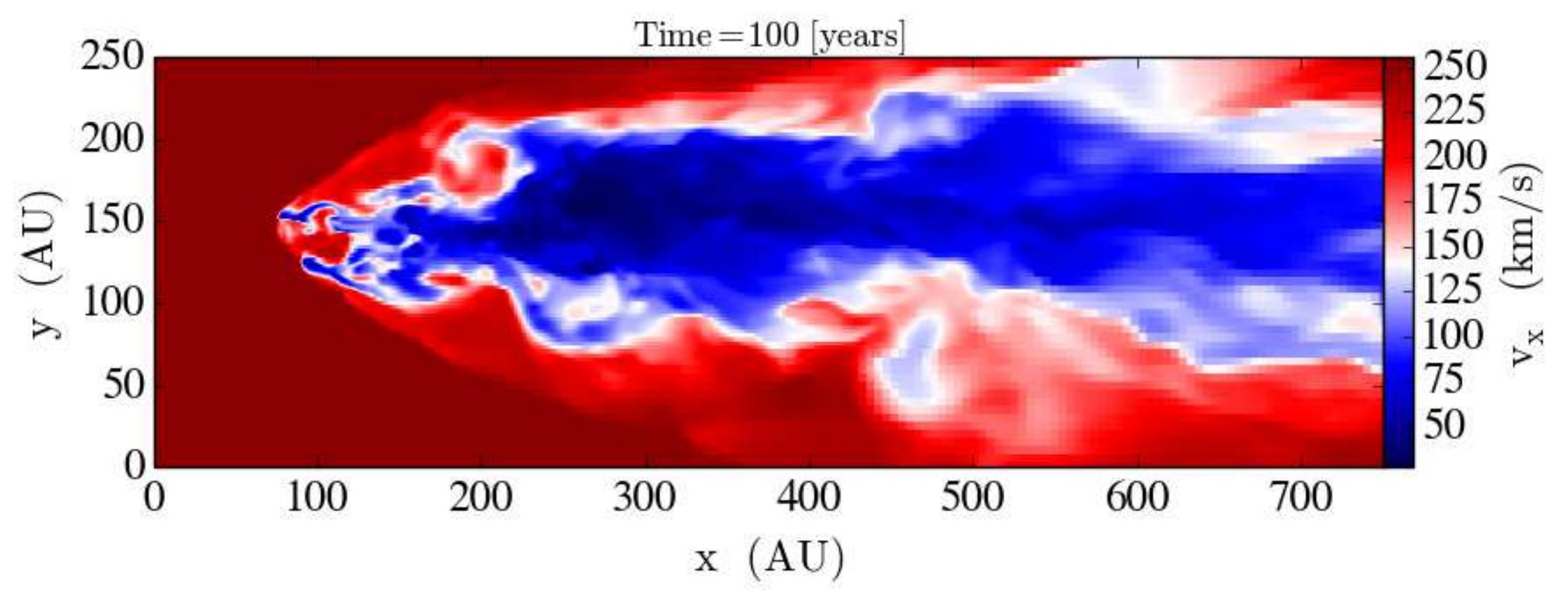}}
\caption{ 
 A slice through the shock model showing the  $x$-velocity 
from the Enzo simulation described in Section 3.3 at t = 100 years. 
The velocities correspond to motions along the line-of-sight.
Owing to the nature of the
simulation, since the bullet is initialized at rest the medium is
flowing past it, small $x$-velocities signify material that has not
been significantly accelerated. Or, when frame-shifted to the case of
a moving bullet, material that has not been significantly
de-accelerated. 
 } 
\label{fig13}
\end{figure*}

\begin{figure*}[htp]
\center{\includegraphics[width=1.0\textwidth,angle=0]
  {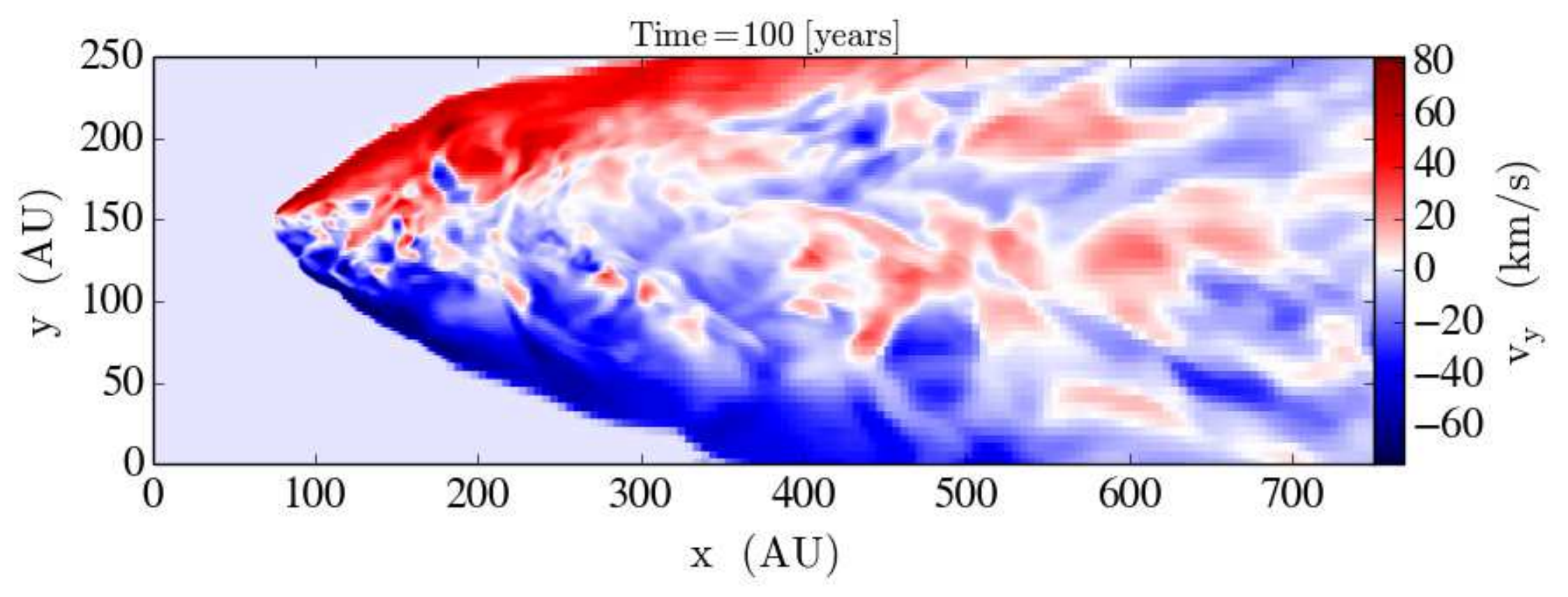}}
\caption{ A slice through the shock model showing the  $y$-velocity 
from the Enzo simulation described in Section 3.3 at t = 100 years. 
The velocities correspond to motions along the line-of-sight.  } 
\label{fig14}
\end{figure*}

\begin{figure*}[htp]
\center{\includegraphics[width=1.0\textwidth,angle=0]
  {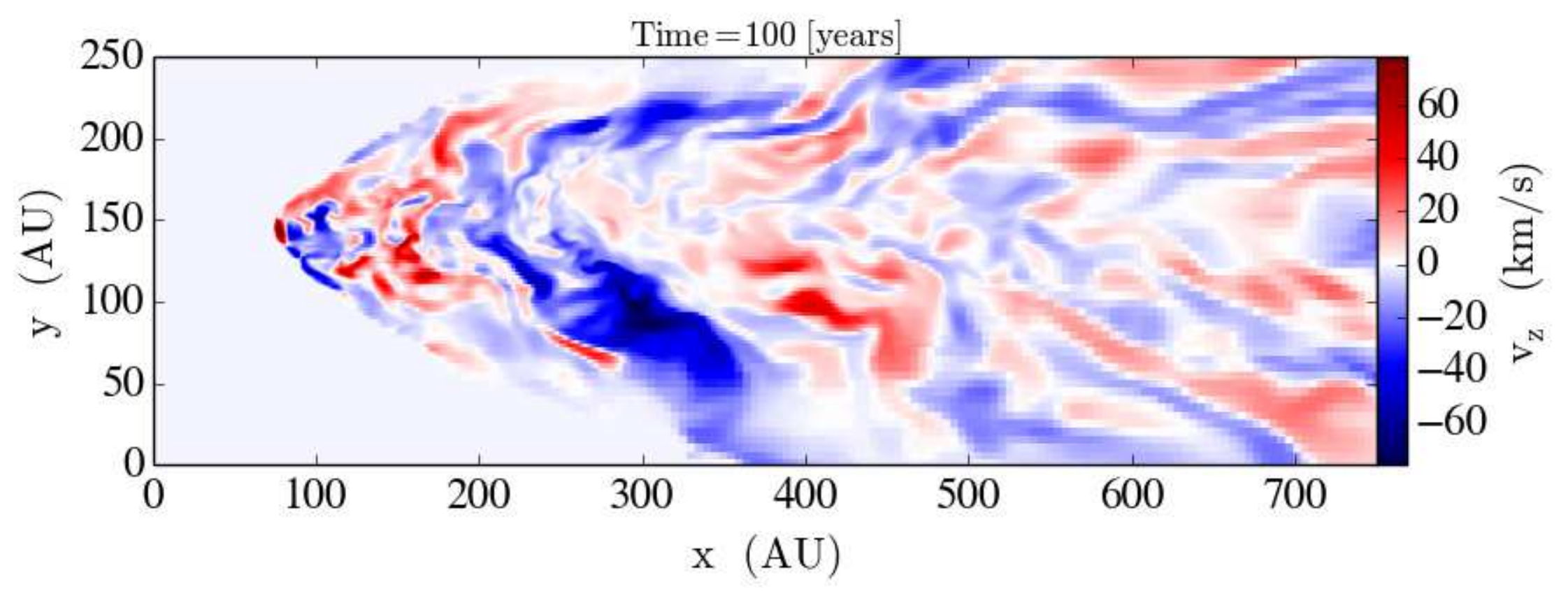}}
\caption{ A slice through the shock model showing the  $z$-velocity 
from the Enzo simulation described in Section 3.3 at t = 100 years. 
The velocities correspond to motions perpendicular to the motion of the bullet
and the line-of-sight.  } 
\label{fig15}
\end{figure*}

%% file: Orion_figures2.tex
\begin{figure*}[htp]
\center{\includegraphics[width=1.0\textwidth,angle=0]
  {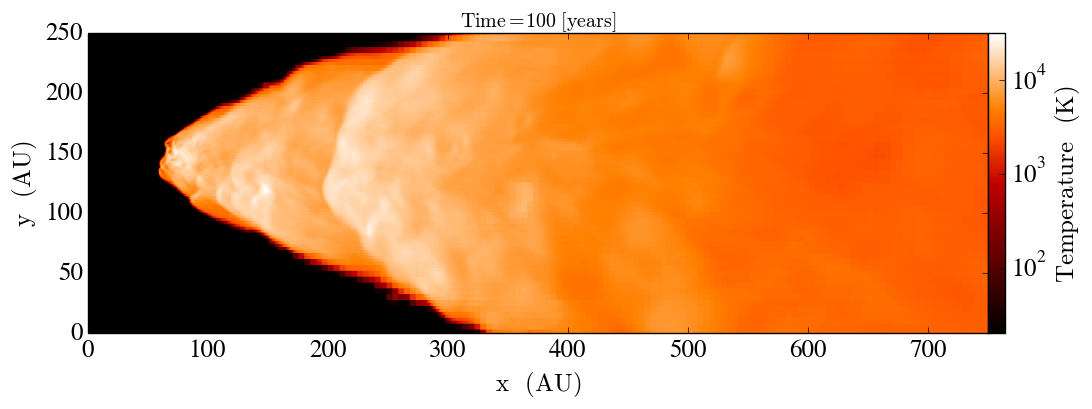}}
\caption{  A density-weighted projection of gas temperature from the Enzo
simulation described in Section 3.3 at t = 100 years. Projected
values are computed such that 
$T_{proj} = \sum \limits_{i} [T_{i} \times \rho_{i}]$ / $\sum \limits_{i} \rho_{i}$, 
where $T_{i}$ is the projected
quantity, $\rho_{i}$ is the weighting quantity, and $i$ denotes the cell
number.   As discussed in the text, actual post-shock temperature in 
slices that are not weighted by density reach temperatures in excess 
of $10^6$ K.} 
\label{fig16}
\end{figure*}

\begin{figure*}[htp]
\center{\includegraphics[width=0.5\textwidth,angle=0]
  {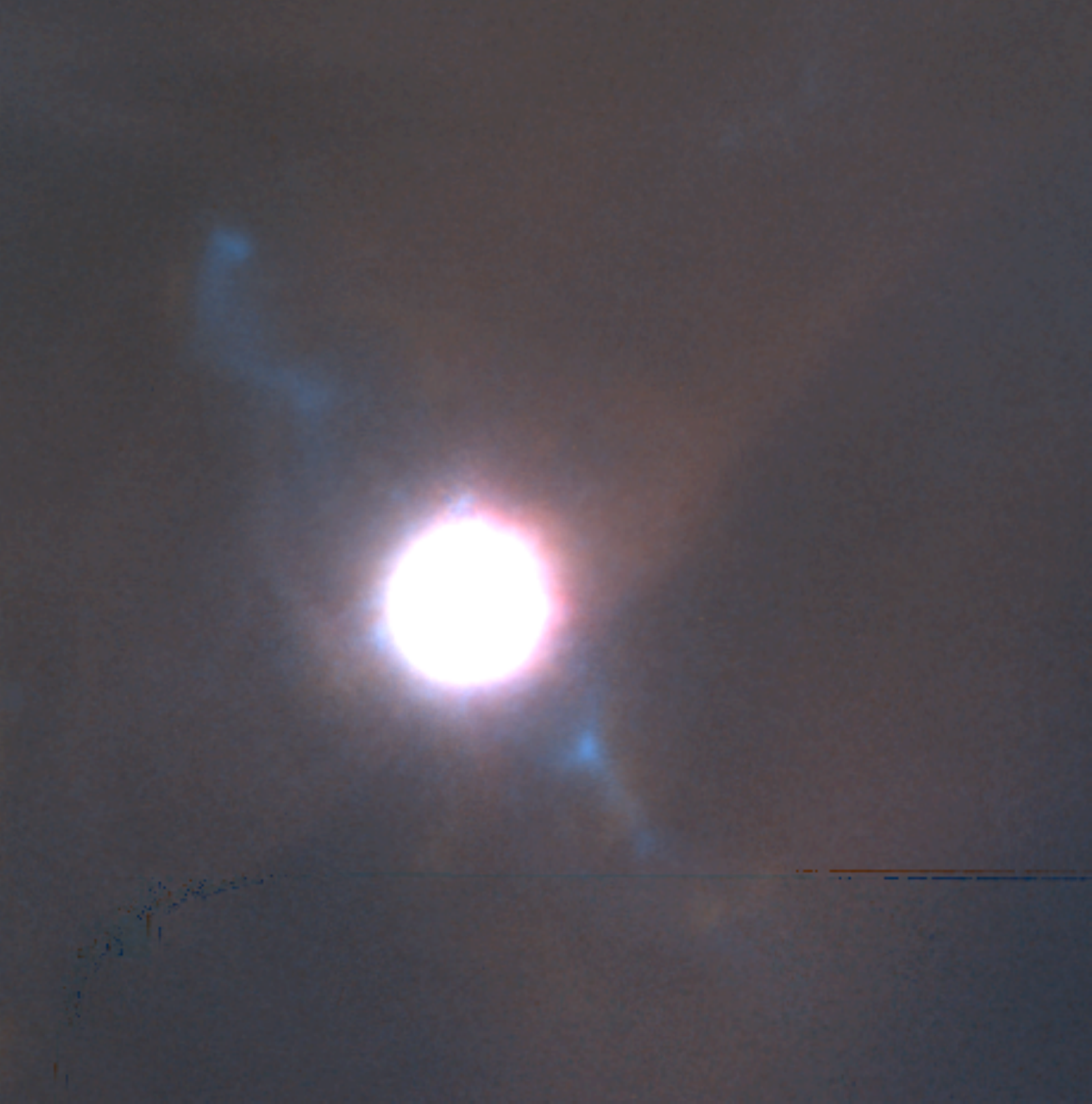}}
\caption{The V2270 \Feii\ jet (blue) and \hh\ emission in the region (orange).} 
\label{fig17}
\end{figure*}

\begin{figure*}[htp]
\center{\includegraphics[width=0.5\textwidth,angle=0]
  {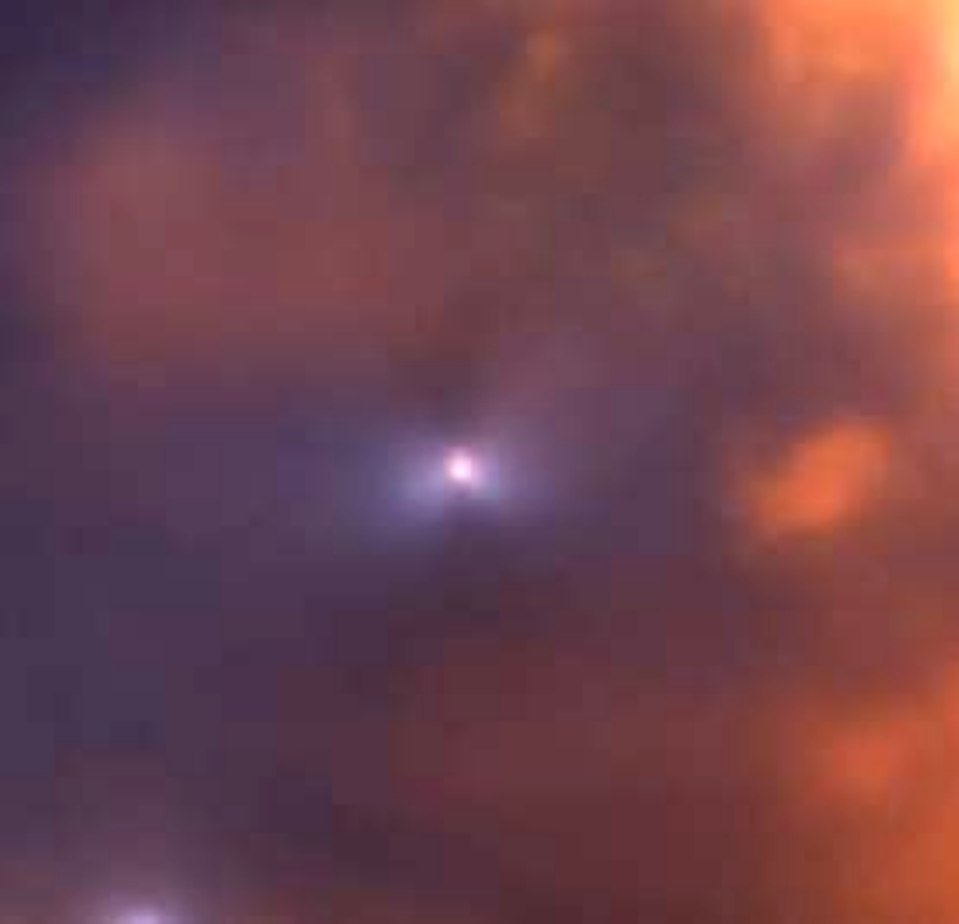}}
\caption{The near-IR disk shadow located 8\arcsec\  east of BN.   \Feii\ emission is shown in
 blue;  \hh\ emission is shown in orange. } 
\label{fig18}
\end{figure*}

%% file: Orion_table1.tex
\begin{table*}[htp]
  
\centering

\caption{~Near-IR Observations of Orion OMC1 
         \label{table1}}

\begin{tabular}{lllllll}
    \textbf{Year  } 			& 
    \textbf{$\alpha$(J2000)}  	& 
    \textbf{$\delta$(J2000)}  	&
    \textbf{Field Size} 		& 
    \textbf{MJD} 	 		&
    \textbf{Comments}    \\
      
\hline
 2007  	&  05:35:12.8  	& -05:20:53  	& 50\arcsec\ $\times$ 50\arcsec\ at 45 \arcdeg\ 	& 54165  	& H$_2$ fingers; Gemini N  \\
 2008 	&  05:35:11.9	& -05:20:47	& 50\arcsec\ $\times$ 50\arcsec\ at 45 \arcdeg\ 	& 54753 	& H$_2$ fingers; Gemini N \\
 2009  	&  05:35:11.9   	& -05:20:47	& 50\arcsec\ $\times$ 50\arcsec\ at 45 \arcdeg\ 	& 55138	& H$_2$ fingers; Gemini N  \\
 2013	&  05:35:13.2	& -05:20:37	& 84\arcsec\ $\times$ 84\arcsec\ at  0\arcdeg\  		& 56290	& OMC1 N;    Gemini S	\\
 2013	&  05:35:12.5	& -05:22:04	& 92\arcsec\ $\times$ 91\arcsec\ at  0\arcdeg\  		& 56323	& OMC1 BN; Gemini S	\\
 2013	&  05:35:17.5	& -05:22:48	& 92\arcsec\ $\times$ 91\arcsec\ at  0\arcdeg\  		& 56323	& OMC1 SE; Gemini S	\\
 
\end{tabular}

\tablecomments{   The \hh  , \Feii  , and K$_s$ images were taken on the same night.}
\end{table*}

%% file: Orion_table2.tex
\begin{table*}[htp]
  
\centering

\caption{~Selected OMC1 bullets and bow shocks 
      \label{table2}}
    
\begin{tabular}{r l l l l r r r  l}
\textbf{Feature} 			& 
\textbf{$\alpha$(J2000)} 	&
\textbf{$\delta$(J2000)}  	& 
\textbf{$\Delta X$}   		&
\textbf{V}   				&
\textbf{PA}   			&
\textbf{D}			&
\textbf{$\tau _{dyn}$}   	&
\textbf{Comments} \\
\textbf{} 				& 
\textbf{(05$^h$ 35$^m$+)}	& 
\textbf{($-$5\arcdeg\ + )}      &
\textbf{(\arcsec)  } 		&
\textbf{(km~s$^{-1}$)} 	&
\textbf{(\arcdeg  )} 		&
\textbf{(\arcsec   )}		&
\textbf{(yrs) } 			&
\textbf{} 				\\
\hline
      
H$_2$ 1 	& 12.625  & 20:55.22  	&  0.88   	& 297	& 346	& 97 		&  643	&  HVCC  H$_2$ knot \\
	    2 	& 12.058  & 20:37.68  	&  0.43   	& 145	& 324	& 116	& 1577	&  Ahead of H$_2$ 1 \\  
	    3	& 12.270	& 20:37.82	&  0.84	& 283	& 344	& 114	&    794	&      " \\
	    4	& 12.351	& 20:38.84	&  0.54	& 182	& 350	& 112	& 1213	&      " \\
	    5	& 12.965	& 21:16.37	&  0.33	& 111	& 338	& 74		& 1314	&   Cluster NE of H$_2$ 1\\
	    6	& 12.989 	& 21:17.73	&  0.53 	& 179	& 347	& 73		&   801	&     " \\
	  S1	& 11.532	& 21:02.24	& 0.30	& 101	& 233	& -		&   -            &  Background flow \\
	  S2	& 11.616	& 20:53.94	& 0.29	& 98		& 227	& -		&  -		&   " \\
	  S3	& 11.614	& 20:54.45	& 0.54	& 182	& 218	& -		& -		&  " \\
	  W1	& 12.543	& 20:52.03	& 0.240	&   80	& -		& -		& 29		& Spreading H$_2$ wake \\
	  W3 & 12.642	& 20:56.45	& 0.138	&  46		& -		& -		& 100	&   "       \\
	  
[Fe II]   1	& 12.521	& 20:50.12	& 0.68	& 229	& 345	& 102	& 863	& HH 207;  Fe II]  bow    \\
	   2	& 12.165	& 20:39.09	& 0.81	& 273	& 345	& 116	& 835	& \\
	   3	& 12.274	& 20:37.96	& 0.79	& 267	& 354	& 114	& 839	& \\
	   4	& 12.353	& 20:38.94	& 0.55	& 186	& 349	& 112	& 1184	& \\
	   5	& 12.898	& 20:34.12	& 0.59	& 199	& 1		& 116	& 1146	& \\
	   6	& 12.932	& 20:34.87	& 0.61	& 206	& 3		& 115	& 1097	& \\
	   7	& 13.066	& 20:35.39	& 0.61	& 206	& 355	& 114	& 1088	& \\
	   8	& 13.052	& 20:39.01	& 0.61	& 206	& 354	& 112	& 1069	& \\
	   9	& 13.010	& 20:40.81	& 0.56	& 189	& 350	& 97	  	& 1009	& \\
	   10	& 13.308	& 20:54.78	& 0.41	& 138	& 354	& 96		& 1367	& \\
	   11	& 13.386	& 20:56.16	& 0.53	& 179	& 353	& 95		& 1043	& \\
	   12	& 13.354	& 20:53.14	& 0.33	& 111	& 357	& 98		& 1736	& \\
	   13	& 12.592	& 20:56.23	& 0.27	& 91		& 354	& 97		& 2096	& \\
	   14	& 12.795	& 21:05.50	& 0.49	& 165	& 354	& 87		& 1036	& \\
	   15	& 11.502	& 21:07.20	& 0.92	& 310	& 338	& 94		& 596	& \\
	   16	& 11.680	& 21:10.86	& 0.71	& 240	& 326	& 89		& 729	& \\
	   17	& 11.792	& 21:12.81	& 0.41	& 138	& 329	& 86		&  1225	&\\
	  
\end{tabular}

\tablecomments{
 {\bf[1]}:   The projected distance from  the suspected point of ejection of
 the stars BN and source I about 500 years ago, D, is measured from  J2000 =
 05:35:14.35, $-$5:22:28.5 (Gomez et al. 2008).  
               }
\end{table*}

\clearpage